# Data Conflict Resolution Using Trust Mappings

Wolfgang Gatterbauer
University of Washington
gatter@cs.washington.edu

Dan Suciu
University of Washington
suciu@cs.washington.edu

## ABSTRACT

In massively collaborative projects such as scientific or community databases, users often need to *agree or disagree* on the content of individual data items. On the other hand, *trust relationships* often exist between users, allowing them to accept or reject other users' beliefs by default. As those trust relationships become complex, however, it becomes difficult to define and compute a consistent snapshot of the conflicting information. Previous solutions to a related problem, the *update reconciliation problem*, are dependent on the order in which the updates are processed and, therefore, do not guarantee a globally consistent snapshot.

This paper proposes the first principled solution to the *automatic conflict resolution problem in a community database*. Our semantics is based on the certain tuples of all stable models of a logic program. While evaluating stable models in general is well known to be hard, even for very simple logic programs, we show that the conflict resolution problem admits a PTIME solution. To the best of our knowledge, ours is the first PTIME algorithm that allows conflict resolution in a principled way. We further discuss extensions to negative beliefs and prove that some of these extensions are hard. This work is done in the context of the *BeliefDB* project at the University of Washington, which focuses on the efficient management of conflicts in community databases.

## 1. INTRODUCTION

In many scientific projects today, a community of users is working together to assemble, revise, and curate a shared data repository. Since the true state of the world is generally not known in many scientific disciplines, users will often have not just overlapping but also conflicting beliefs about the world. Hence, as the community accumulates knowledge and the database content evolves over time, it will inevitably contain conflicting information. An example domain with highly disputed and still changing beliefs is the current state of knowledge on the Indus script [17].

|   | *glyph* | *origin* |   |
|---|---|---|---|
| $b_1$ | ᛰ | ship hull | Alice |
| $b_2$ | ᛰ | cow | Bob |
| $b_3$ | ᛰ | jar | Charlie |
| $b_4$ | ᚴ | fish | Bob |
| $b_5$ | ᚴ | knot | Charlie |
| $b_6$ | ↑ | arrow | Bob, Charlie |

(a)

| *glyph* | *origin* |
|---|---|
| ᛰ | ship hull |
| ᚴ | fish |
| ↑ | arrow |

(b)

**Figure 1:** Origins of three Indus glyphs as asserted by archeologists Alice, Bob and Charlie (a). Alice's beliefs after applying trust mappings from Fig. 2 (b).

EXAMPLE 1.1 (INDUS SCRIPT). *The Indus civilization flourished about 2600 to 1900 B.C. in what is now eastern Pakistan and northwestern India. No historical information exists about the civilization, but archaeologists have uncovered samples of their writing on stamp seals, sealings, amulets, and small tablets. To this day, the script on these objects remains undeciphered and various claims of decipherments have been put forward. One important step in such interpretations is determining the "origin" of the glyph which is the actual motif that was stylized into the glyph. For example, the origin of the glyph* ᛰ *has been attributed by different archeologists[1] to be either a* ship hull, *a* cow, *or a* jar. *Similarly, the glyph* ᚴ *has been interpreted to originate from a* fish *or a* knot, *whereas the glyph* ↑ *is widely agreed to represent an* arrow *[15]. The current state of knowledge can be represented in a relational table (Fig. 1a) where conflicting beliefs of researchers (here simplified with Alice, Bob and Charlie) are represented by tuples (here tuples $b_1$ to $b_6$) with partial key violations, highlighting the lack of agreement among archeologists.*

Recent work has proposed *trust mappings* between users in order to support collaboration and data sharing [19]. A trust mapping is a statement that a user is willing to accept another user's data value. Priorities are further used to specify how to resolve conflicts between data values coming from different trusted users. Figure 2 illustrates three trust mappings, two defined by Alice ($m_1$ and $m_2$) and one defined by Bob ($m_3$). Figure 1b shows Alice's version of the data after applying these mappings to Example 1.1: The second and third tuples result from her trust of Bob and Charlie. Where they disagree (on ᚴ), Alice sees Bob's value (fish) instead of Charlie's value (`knot`) because she assigned to Bob a higher priority (100) than to Charlie (50)[2].

---

[1] Parpola, Mahadevan, and Knorozov [15]

Several systems have adopted some form of conflict handling or trust mapping in order to facilitate data sharing among users [7, 9, 11, 16, 19] (see Fig. 3 for a comparison of their features). However, providing a consistent semantics to a set of trust mappings is a challenging problem. The current state of the art is the approach taken by the Orchestra system in the context of *update exchange* [9]. Updates are treated one at a time, in a *First-In First-Out* manner. When an update is published by a user, the new value is propagated according to the trust mappings to other users, and conflicts are resolved based on the priorities. At each moment, a user is provided with a snapshot of the current database, according to other users' beliefs and her own trust relationships. However, this snapshot may be inconsistent because (*i*) it depends on the order in which the updates are processed, and (*ii*) it may become inconsistent as a result of other updates.

EXAMPLE 1.2 (INDUS SCRIPT CONTINUED). *Consider the following sequence of updates starting with an empty database:*

| Time | User | Update for $\mho$ | Beliefs for $\mho$ | | |
|---|---|---|---|---|---|
| | | | Alice | Bob | Charlie |
| 0 | | | - | - | - |
| 1 | Charlie | insert jar | - | - | jar |
| 2 | | | jar | - | jar |
| 3 | | | jar | jar | jar |
| 4 | Bob | insert cow | ? | cow | jar |

*Charlie is the first user to insert the origin of $\mho$ as* `jar`, *which propagates to Alice and Bob. When Bob inserts* `cow`, *the system fails to propagate this to Alice because she already has a data value acquired at an earlier timestamp. Thus, Alice continues to see* `jar`. *This is inconsistent, because Alice trusts Bob more than Charlie. Had the updates been performed in reverse order (first Bob inserts* `cow` *then Charlie inserts* `jar`) *then Alice would end up with* `cow`. *Thus, the snapshot for Alice depends on the order in which the updates are performed, which is undesirable. One may attempt to address that by storing the lineage of each data value: the system could then update Alice's value because it knows it came from Charlie. However, maintaining the lineage of all values in a large system is difficult, because an update to one value may affect the lineage of many other data values.*

*A second problem is that update propagation cannot handle updates or* revocation *of data values. Consider the following sequence, again starting from an empty database:*

| Time | User | Update for $\mho$ | Belief for $\mho$ | | |
|---|---|---|---|---|---|
| | | | Alice | Bob | Charlie |
| 0 | | | - | - | - |
| 1 | Charlie | insert jar | - | - | jar |
| 2 | | | jar | - | jar |
| 3 | | | jar | jar | jar |
| 4 | Charlie | update jar → cow | ? | ? | cow |

*Both Alice and Bob import* `jar` *from Charlie, and both should update their belief when Charlie updates his to* `cow`. *How-*

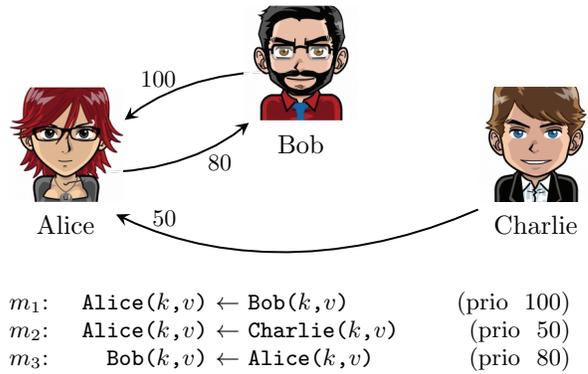

$m_1$:    `Alice`$(k,v) \leftarrow$ `Bob`$(k,v)$      (prio 100)
$m_2$:    `Alice`$(k,v) \leftarrow$ `Charlie`$(k,v)$      (prio 50)
$m_3$:    `Bob`$(k,v) \leftarrow$ `Alice`$(k,v)$      (prio 80)

**Figure 2: Example trust mappings between archeologists Alice, Bob and Charlie.**

*ever, this is difficult to achieve even if one keeps the lineage of each value. The problem is that Alice and Bob trust each other with highest priority: Alice keeps her* `jar` *value because of Bob' value* `jar`; *Bob keeps his value because of Alice's value.*

In this paper we propose an efficient and principled solution to the *conflict resolution problem*: given a network of priority trust mappings, find the values that are believed (and trusted) by each user. Our first contribution is a definition of conflict resolution based on a *stable solution*. A stable solution is simply a global assignment of values to users such that all trust mappings are satisfied, and that each value has a lineage to a value that was inserted explicitly by some user. We show that stable solutions correspond precisely to the *stable models of logic programs with negation*. However, computing stable models is notoriously hard: even if each rule of the program has a single predicate which is negated, deciding if a stable model exists for a given logic program is NP-hard [4, pp. 396]. We show that off-the-shelf logic program solvers scale exponentially when applied to conflict resolution. Instead, we describe a new, efficient algorithm that runs in *quadratic time in the number of trust mappings*, and computes for each user both the possible values and the certain values. To the best of our knowledge, this is the first solution to the conflict resolution problem that is both *principled and efficient*. We describe further extensions to answer queries about the conflicts themselves, such as finding pairs of users that disagree in at least one stable solution, or finding the *lineage of a belief*.

Secondly, we study the impact of *constraints* on the conflict resolution problem. Attribute-value constraints are defined by users, and can be, for example, a range-constraint for a numerical attribute, or an inclusion constraint (checking that the value appears in a reference database) for a categorical attribute. Once a user defines a constraint, the effect is that she rejects all values that do not satisfy the constraint, hence we also refer to a constraint as a set of *negative beliefs*.

The simplest way to handle constraints is to restrict their usage locally: the constraint is only used to decide whether to accept or reject a value coming from another trusted user, and the constraint itself is not further propagated to other users. Only the data values are propagated. We call this the *agnostic* paradigm. Another reasonable approach is to treat constraints and data values equally and to propagate them together, based on the prioritized trust mappings. Thus, if Alice trusts Bob, then she also trusts Bob's constraints, and therefore may reject a value from Charlie if that value violates Bob's constraint. We call this paradigm *eclectic*. Surprisingly, we show that both the agnostic and the eclectic paradigms are *computationally hard*: computing the possible values is NP-hard, and computing the certain values is co-NP hard. Therefore, we propose a third paradigm of dealing with constraints, called *skeptic*. Here constraints continue to be propagated, but, in addition, each data value $v$ is augmented with a constraint that rejects all other data values except $v$. When $v$ is accepted, the constraint is redundant, but its role becomes important later downstream, if $v$ is rejected (e.g. because of some other user's constraint). We show that the skeptic paradigm to conflict resolution can be computed efficiently (in quadratic time in the number of trust mappings). Thus, we propose the skeptic paradigm as our solution to handle constraints during conflict resolution.

---

[2]Note that priorities are assigned by users and only serve to impose *a total preorder on their trusted parties*. Priorities of mappings assigned by different users, e.g. $m_2$ with priority 50 and $m_3$ with priority 80, cannot be compared.

| | conflicts | trust mapping | priorities | update independ. | revokes | cycles | consensus queries |
|---|---|---|---|---|---|---|---|
| Orchestra [9, 19] | x | x | x | | | x | |
| FICSR [16] | x | | | | | | |
| BeliefDB [7] | x | | | x | x | | x |
| Youtopia [11] | | x | | x | x | | |

**Figure 3: Recently proposed systems that model conflicts or data sharing for a community of users.**

Finally, we study *bulk conflict resolution*, when similar trust relationships are applied to a large number of data objects. Here, the complexity is dominated by a large number of objects, while the number of users is assumed to be small. We show that by simple changes to our two algorithms (without and with constraints) we can translate the trust mappings into SQL and execute them in a standard relational database.

In summary, our contributions with this paper are:
- We define a principled solution to the conflict resolution problem, called *stable solution*, and give an algorithm that runs in quadratic worst time (Sect. 2).
- We describe three paradigms for handling constraints in conflict resolution and prove that two are hard. For the third one, we give an algorithm that runs in quadratic worst time (Sect. 3).
- We discuss conflict resolution in bulk and describe a solution based on a translation into SQL (Sect. 4).
- We conduct experiments that show that, while quadratic in the the worst case, our proposed solutions actually scale linearly in most cases (Sect. 5).

## 2. BASIC CONFLICT RESOLUTION

In this section, we define our model for handling conflicts through priority trust mappings. We do not consider constraints here, but discuss those later in Sect. 3.

The database consists of a collection of objects. We assume w.l.o.g. that each object is identified by a key $k$ and has a single attribute. We write $(k, v)$ when the object with identifier $k$ has attribute value $v$. Given an object, different users may have conflicting beliefs about the correct value of its attribute. Thus, the attribute may take one of several values $v_1, v_2, \ldots$ according to different users' beliefs. We denote $D$ the set of possible data values, and $U$ the set of users. Thus, user $x \in U$ may believe that $(k, v)$ is correct, user $y$ may believe $(k, w)$ is correct, where $v, w \in D$, while user $z$ may have no opinion about the value of the object $k$.

Users' beliefs may differ from object to object. In this and the following section, we treat each object separately and therefore omit mentioning $k$ altogether. In Sect. 4, we discuss how to efficiently handle multiple objects.

Thus, a user either has an *explicit belief* about the value of the object, or can derive her belief *implicitly* through *priority trust mappings*. We define these notions formally.

DEFINITION 2.1 (EXPLICIT BELIEF $b_0$). *An explicit belief is a partial function from users to values with $b_0(x)$ being the value believed by user $x$.*

DEFINITION 2.2 (PRIORITY TRUST MAPPING $m$). *A priority trust mapping (or trust mapping, in short) is a triple $m = (z, p, x)$, where $z, x$ are users and $p$ is an integer. The meaning is that user $x$ trusts the value from user $z$ with priority $p$. We call $z$ a parent of $x$, and $x$ a child of $z$.*

DEFINITION 2.3 (PRIORITY TRUST NETWORK TN). *A Priority Trust Network (or trust network, in short) is a labeled graph $TN = (U, E, b_0)$, where $U$ is a set of users, $E$ is a set of priority trust mappings, and $b_0$ is an explicit belief.*

Given a TN, each user $x$ computes her belief $b(x)$ as follows: If $x$ has an explicit belief $b_0(x)$ then this is also her belief. Otherwise, $x$ computes her belief *implicitly* by choosing a trust mapping $(z, p, x)$ with highest priority (i.e. largest $p$ with ties broken arbitrarily), such that the parent $z$ has some belief, and defines $b(x) = b(z)$. Thus, the explicit beliefs stated by some of the users propagate through the graph to other users. If a user has two trusted parents with different beliefs, then the conflict is resolved according to the priority, with ties broken arbitrarily.

DEFINITION 2.4 (STABLE SOLUTION $b$). *A stable solution of a trust network $(U, E, b_0)$ is a partial function $b$ from users to values, so that for every user $x$ with $b$ defined, there exists a path $x_0 \to x_1 \to \ldots \to x_n = x$, s.t.:*
(1) $x_0$ *has an explicit belief* $b_0(x_0)$,
(2) *all nodes along the path have the same implicit belief* $b(x_i) = b_0(x_0)$, *and*
(3) *at no node $x_i$ there is a trust mapping $(x'_{i-1}, p', x_i)$ with higher priority $p' > p$ than $(x_{i-1}, p, x_i)$ and with a conflicting belief $b(x'_{i-1}) \neq b(x_i)$ at parent $x'_{i-1}$.*

*In this case, we call the sequence $x_0, \ldots, x_n$ the* lineage *of the belief $b(x)$. In other words, every belief $b(x)$ can be traced to some explicit belief along paths that are not dominated with conflicting values. $b(x)$ remains undefined only if no parent of $x$ has a belief and $x$ has no explicit belief either.*

*Resolving a trust network* can be thought of as a *conflict resolution* process: users that have explicit and distinct beliefs conflict with each other, and users without explicit beliefs decide whom to trust based on the priority trust mappings. We illustrate with two examples:

EXAMPLE 2.5 (SIMPLE TN). *Fig. 4a shows a simple network that has a single stable solution: $b(x_1) = v$ (and $b(x_2) =$*

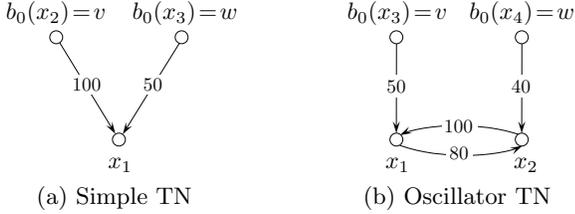

**Figure 4:** Two example trust networks with either one (a) or two (b) stable solutions.

$v$, $b(x_3) = w$). Consider now Fig. 2 and assume the single explicit belief $b_0(\texttt{Charlie}) = \texttt{jar}$. Then the unique solution is $b(\texttt{Alice}) = b(\texttt{Bob}) = \texttt{jar}$. On the other hand, if both Charlie and Bob have explicit beliefs $b_0(\texttt{Charlie}) = \texttt{jar}$, $b_0(\texttt{Bob}) = \texttt{cow}$, then the unique solution is $b(\texttt{Alice}) = \texttt{cow}$.

EXAMPLE 2.6 (OSCILLATOR TN). *Now consider the trust network in Fig. 4b. It has two stable solutions, one with $b(x_1) = b(x_2) = v$, the other with $b(x_1) = b(x_2) = w$. To see that the first is a stable solution, note that $x_1$ and $x_2$ derive their beliefs from $x_2$ and $x_1$ respectively (cyclic), and both have a lineage that can be traced back to $b_0(x_3)$. On the other hand, setting $b(x_1) = b(x_2) = u$, where $u$ is an arbitrary value, is not a stable solution, because $u$ doesn't have a lineage to an explicit belief.*

The last example illustrates an important issue in solving a trust network: if the network has *cycles*, then there can be *more than one stable solution*, even if for each node, the trust mappings impose a total order on its parents. Cycles in trust mappings occur naturally since mappings are defined by individual users, and those often form groups that mutually trust each other more than others. Therefore, a conflict resolution system must cope with cycles, and this is a difficult problem as we have seen in Example 1.2.

## 2.1 Problem Statements

We define a principled approach to conflict resolutions in terms of *certain beliefs*. For technical purposes, we define them together with their duals, the *possible beliefs*:

DEFINITION 2.7 (CERTAIN / POSSIBLE BELIEF). *Let $x$ be a user and $v$ be a value. We say that $v$ is a* certain belief *if for every stable solution $b$, $b(x) = v$. We say that $v$ is a* possible belief *if there exists a stable solution $b$ s.t. $b(x) = v$.*

The main problem that we study in this paper is *for each user $x$, find their certain beliefs*, denoted $\mathsf{cert}(x)$. We call it *resolving a trust network*. For a simple illustration, the certain beliefs in Example 2.6 are: $\mathsf{cert}(x_1) = u$, $\mathsf{cert}(x_2) = v$, $\mathsf{cert}(x_3) = \mathsf{cert}(x_4) = \emptyset$. Each user's certain belief represent a snapshot of the conflicting information in the network. If the network has a unique stable solution $b$, then each user $x$ with at least one parent with a belief, also has a well defined certain belief, and $\mathsf{cert}(x) = b(x)$; if there is more than one solution, then for some of those users $\mathsf{cert}(x) = \emptyset$.

In addition, we show that our techniques can be extended to solve several other related problems, such as:
- *Agreement checking:* Find pairs of users $x, y$ who agree in all stable solutions $b$: $b(x) = b(y)$.
- *Consensus value:* Given two users $x, y$, find all values $v$ such that in every stable solution $b$, $b(x) = v$ iff $b(y) = v$.
- *Lineage computation:* Given a user $x$ and possible value $v$ for $x$, compute a lineage for $v$.

## 2.2 Binary Trust Networks

A *Binary Trust Network* (BTN) is a TN where every node $x$ has at most two incoming edges and the explicit beliefs $b_0(x)$ are defined only for *root nodes* $x$ (nodes without parents but with explicit beliefs). We further assume that every node in a BTN is reachable by a path from some root node. If $x$ is not reachable, then $b(x)$ is undefined in any stable solution and may be safely removed from the BTN. The networks in Fig. 2 and Fig. 4 are binary.

If $x$ has a single parent $z_1$, or two parents $z_1, z_2$ with priorities $p_1 > p_2$, then $z_1$ is called *preferred parent*. Otherwise, it is called a *non-preferred parent*. Note that if there are two parents with the same priority, they are both non-preferred.

PROPOSITION 2.8. (BINARY TRUST NETWORK EQUIVALENCE) *Every trust network TN is equivalent to a binary trust network BTN of maximum double size, where size is the number of trust mappings.*

We prove Proposition 2.8 in the appendix and from now on, and w.l.o.g., consider only binary trust networks.

## 2.3 Logic Programming and Stable Models

One possible approach to solve a trust network is to use Logic Programming (LP). Techniques for LP have been studied extensively in the literature [4] and there are tools for LP solving. In this section, we explore the applicability of LP to solving a trust network. We associate the following LP to a BTN. There is a unary predicate $U_x$ for each user $x$, and a unary predicate $C_{x,z}$ for every non-preferred edge $z \to x$. If $x$ has an explicit belief $v$, then $U_x$ is an EDB predicate, satisfying the single predicate $U_x(v)$; otherwise it is an IDB predicate. All $C_{x,z}$ predicates are IDB predicates. The rules in LP are the following. For every preferred edge $y \to x$ we have rule (1) below, and for every non-preferred edge $z \to x$ we have rules (2a), (2b) below:

$$U_x(r) \leftarrow U_y(r) \qquad (1)$$
$$C_{x,z}(s) \leftarrow U_z(s), U_x(r), r \neq s \qquad (2a)$$
$$U_x(s) \leftarrow U_z(s), \neg C_{x,z}(s) \qquad (2b)$$

The first rule says that $x$ should believe everything that its preferred parent $y$ does. The second and third rules state that $x$ should believe $z$, except when it conflicts with its own belief (presumably obtained from the preferred parent, or from the other non-preferred parent). The intermediate predicate $C_{x,z}$ is introduced for technical reasons, to guarantee *safety* of the resulting LP.

The semantics of a LP is given in terms of a *stable model*. We briefly review the definition and refer to [4] for details. Consider the grounded logic program $P$, obtained by grounding each rule with all values in the active domain: each rule in the grounded LP has some positive and some negative predicates (In the case of BTN's, there is at most one negative predicate). Consider a model $M$. The *reduct* of $P$ by $M$ is the logic program $P^M$ obtained by (a) removing all rules $P$ where some negative predicate is false in $M$, and (b) removing all negative predicates from the remaining rules. Note that the reduct has no negations. The model $M$ is called a *stable model* if it is the minimal fixpoint model of the reduct $P^M$. A ground fact is called *certain* if it belongs

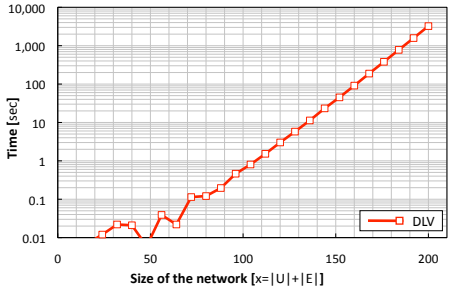

**Figure 5:** Resolving a trust network with LP solvers is exponential in the size of the network.

---

**Algorithm 1:** Resolution Algorithm

**Input**: $BTN = (U, E, b_0)$
**Output**: $\mathsf{poss}(x)$ for each node $x \in U$

I  $\mathsf{closed} \leftarrow \emptyset$
  **foreach** node $x$ with $b_0(x)$ defined **do**
  $\quad \mathsf{poss}(x) \leftarrow \{b_0(x)\}$
  $\quad$ add $x$ to $\mathsf{closed}$

  $\mathsf{open} \leftarrow U - \mathsf{closed}$
M **while** $\mathsf{open} \neq \emptyset$ **do**
S1 $\quad$ **if** $\exists$ preferred edge $z \to x$ with $z \in \mathsf{closed} \wedge x \in \mathsf{open}$ **then**
  $\quad\quad \mathsf{poss}(x) \leftarrow \mathsf{poss}(z)$
  $\quad\quad$ move $x$ from $\mathsf{open}$ to $\mathsf{closed}$
S2 $\quad$ **else**
  $\quad\quad$ Let $SCC(\mathsf{open})$ be the SCC graph constructed from the open nodes. Let $S$ be a minimal SCC. Let $\mathsf{possS} \leftarrow \emptyset$.
  $\quad\quad$ **foreach** edge $z \to x$ with $z \in \mathsf{closed} \wedge x \in S$ **do**
  $\quad\quad\quad$ add all values from $\mathsf{poss}(z)$ to $\mathsf{possS}$
  $\quad\quad$ **foreach** $x \in S$ **do**
  $\quad\quad\quad \mathsf{poss}(x) \leftarrow \mathsf{possS}$
  $\quad\quad$ move all nodes of $S$ from $\mathsf{open}$ to $\mathsf{closed}$

---

to all stable models; it is called *possible* if it belongs to some stable model. We prove the following in Appendix B.

THEOREM 2.9 (LOGIC PROGRAM EQUIVALENCE). *Every stable solution to a BTN is a stable model to the associated LP, and vice versa.*

EXAMPLE 2.10 (OSCILLATOR). *The LP corresponding to the oscillator from Example 2.6 is:*

$U_3('v') \leftarrow \qquad\qquad U_4('w') \leftarrow$
$U_1(r) \leftarrow U_2(r) \qquad\qquad U_2(r) \leftarrow U_1(r)$
$C_{1,3}(s) \leftarrow U_3(s), U_1(r), r \neq s \qquad C_{2,4}(s) \leftarrow U_4(s), U_2(r), r \neq s$
$U_1(s) \leftarrow U_3(s), \neg C_{1,3}(s) \qquad U_2(s) \leftarrow U_4(s), \neg C_{2,4}(s)$

*It has two stable models, $M_1 = \{U_1('v'), U_2('v'), U_3('v'), U_4('w')\}$, and $M_2 = \{U_1('w'), U_2('w'), U_3('v'), U_4('w')\}$.*

In summary, one can resolve a trust network as follows. Rewrite the BTN into a logic program, then use a LP solver to compute the certain tuples. Unfortunately, solving logic programs is hard, even in the most restricted settings. Deciding whether a logic program has at least one stable model is coNP-hard even if all rules have a single atom in the body and that atom is negated; deciding if a ground tuple is a certain tuple is coNP-hard [13]. These theoretical results translate into running times that increase exponentially in the size of the BTN. Figure 5 shows the running time of the state-of-the-art LP-solver DLV [12] on increasingly larger BTNs composed of several oscillators from Fig. 4b. One can see the clear exponential trend: the LP solver becomes impractical on graphs larger than 150 nodes. (We report more extensive experiments in Sect. 5.) Thus, relying on a general purpose LP solver is not a practical solution. Instead, we describe in the next section a new algorithm for resolving a trust network which runs in quadratic time of its size.

## 2.4 An Algorithm for Conflict Resolution

Algorithm 1 takes as input a BTN and computes the set of possible values $\mathsf{poss}(x)$ for every node $x$. The sets of certain values $\mathsf{cert}(x)$ can then be derived using the following rules: $\mathsf{cert}(x) = \{a\}$ if $\mathsf{poss}(x) = \{a\}$, and $\mathsf{cert}(x) = \emptyset$ otherwise.

The algorithm maintains a set $\mathsf{closed}$ of all nodes $x$ for which $\mathsf{poss}(x)$ is already computed. This set is initialized with all *root nodes*, i.e. all nodes with explicit beliefs (I). $\mathsf{open}$ contains all other nodes. During the main loop (M), the algorithm performs two steps: Step 1 (S1) propagates greedily $\mathsf{poss}(x)$ along preferred edges, closing the destination nodes. Step 2 (S2) handles the case when no more preferred edges can be traversed, and we describe it next:

A *strongly connected component* (SCC) in a graph is a set $S$ of nodes so that there exists a path from $x$ to $y$ and a path from $y$ to $x$, for every two nodes $x, y \in S$. The *SCC graph* is formed by contracting the vertices of each SCC. It is known that this graph is acyclic and can be calculated in $\mathcal{O}(n)$ time using Tarjan's algorithm [18]. In the second step (S2), the algorithm computes the SCC graph of $\mathsf{open}$ nodes, and chooses a component $S$ that is *minimal* in the SCC graph. In other words, $S$ is an SCC that has no incoming edges from other SCCs. $S$ may still have incoming edges, but they are all coming from $\mathsf{closed}$, and moreover, they are all non-preferred edges. The algorithm defines $\mathsf{poss}(x)$, for all nodes in $S$ as the union of all possible values of all parents of nodes in $S$. It then closes all nodes in $S$.

EXAMPLE 2.11 (OSCILLATOR). *We illustrate the algorithm for Example 2.6. Initially, $\mathsf{closed} = \{x_3, x_4\}$, and $\mathsf{poss}(x_3) = \{v\}$, $\mathsf{poss}(x_4) = \{w\}$. There are no preferred edges that can be traversed, so the algorithm proceeds with the second step, computing the connected components of $\mathsf{open}$. There is a single component, $\{x_1, x_2\}$, and their possible values are set to $\mathsf{poss}(x_3) \cup \mathsf{poss}(x_4) = \{v, w\}$. The output of the algorithm is: $\mathsf{cert}(x_3) = \{v\}$, $\mathsf{cert}(x_4) = \{w\}$, and $\mathsf{cert}(x_1) = \mathsf{cert}(x_4) = \emptyset$.*

THEOREM 2.12 (RESOLUTION ALGORITHM). *Let $n$ be the number of nodes in a BTN. Algorithm 1 runs in time $\mathcal{O}(n^2)$ and correctly computes the set of possible tuples for all nodes $x$ in BTN.*

The running time follows from the fact that the SCC graph can be computed in time $\mathcal{O}(n)$. Note that the runnig time is not $\mathcal{O}(n)$, as the SCC's may need to be re-calculated at each iteration as the set $\mathsf{open}$ changes at each iteration: we show in the technical report that the algorithm takes $\Omega(n^2)$ in the worst case. We prove correctness in Appendix A.

## 2.5 Discussion and Extensions

Let's step back and examine what we have achieved. We have fixed an object $k$, and considered a priority trust network. Our goal is to give each user a snapshot of the data that is consistent with the entire network. Definition 2.4 defines a stable solution for this network, but in general there may be several stable solutions. We have proposed the *certain* values as the snapshot to be shown to the user, and described Algorithm 1, which computes the certain values (and the possible values too) in time quadratic in the number of users. The algorithm may need to be run separately for each object $k$, an issue that we will address in Sect. 4.

The important property of our approach is that both the definition and the algorithm are *order-invariant*: they do not rely on any order in which conflicts are to be resolved. The result is a consistent snapshot of the conflicting information. By contrast, as we have seen, prior approaches to conflict resolution process the explicit beliefs in a fixed order (e.g. in the order of their transaction time), and the result depends on this order. As a consequence, if any explicit belief is updated, e.g. some belief is revoked, there may be no way to re-compute a consistent snapshot. In our approach, if an explicit belief is updated, we will simply re-run the algorithm and obtain another consistent snapshot.

As a further benefit of a principled approach, we mention here two extensions of our algorithm that allow the system to answer more complex queries, as those mentioned in Sect. 2.1.

**Retrieving lineage.** We show how to extend the algorithm to compute the lineage of each possible value. Whenever we insert a value $v$ into $poss(x)$, store a pointer back to the value $v \in poss(z)$ that produced this possible value at $x$: for Step 1 this is a value in the preferred parent, for Step 2 there can be several (user, value) pairs from outside the set $S$: store pointers to all of them. Thus, from each value $v \in poss(x)$ we can trace back several lineages. Note that this method is not complete: Step 1 misses some lineages that come to $x$ via non-preferred edges. However, it has the property that each possible value has at least one lineage that the system can return to the user.

**Pairs of possible values.** For any two users $x, y$ in a binary trust network, denote:

$poss(x, y) = \{(v, w) \mid \exists$ stable solution $b: b(x) = v, b(y) = w\}$

Thus, $poss(x, y)$ denotes the set of pairs of values that $x$ and $y$ can take *together*. Note that if $(v, w) \in poss(x, y)$ then $x \in poss(x)$ and $y \in poss(y)$, but the converse is not true. For example in Fig. 4b, $poss(x_1, x_2)$ contains the pairs $(v, v)$ and $(w, w)$, but not $(v, w)$ or $(w, v)$.

PROPOSITION 2.13 (POSSIBLE PAIRS). *Algorithm 1 can be extended to compute* $poss(x, y)$ *for all pairs of users $x, y$. The modified algorithm runs in time $\mathcal{O}(n^4)$ where $n$ is the number of users.*

The sets $poss(x, y)$ allow us to go beyond the snapshot consisting of certain tuples, and answer more complex queries about the conflicts and the reconciliation. For example, the *agreement checking* query mentioned in Sect. 2.1 can be answered as $\{(x, y) \mid \forall (v, w) \in poss(x, y) \Rightarrow v = w\}$.

## 3. CONFLICT RESOLUTION WITH CONSTRAINTS

In this section, we extend our approach to constraints which we model as *negative beliefs*. So far, we have only considered positive beliefs, i.e. a user either believes that the value of an object is $v$ or has no opinion at all. A negative belief, in contrast, states that the value of the object is *not* $v$. We denote with $(k, v)+$ a positive belief, and with $(k, v)-$ a negative belief.

Constraints occur naturally in collaborative systems and enable users to filter the data values they accept. For example, one users may define the constraint *the value of the 'carbon-date' attribute is between 1,200 and 40,000*: this corresponds to a negative belief for every value $v$ outside the range. Or, another user may rely on a reference database before accepting a value, e.g. *the value of the 'translation attribute' must be in the 'list-of-known-words'*. These constraints are used to refuse a value from a trusted user and therefore affect the global conflict reconciliation. In addition, a user may state a negative belief explicitly in order to refute another user's statement. For example, user Alice may state that the origin of ꙮ is cow, written $(k_1, \text{cow})+$. User Bob may disagree. He does not know what the origin of the glyph is, but believes it cannot be a cow. His belief is thus $(k_1, \text{cow})-$. Bob may accept other values, such as horse or jar, from users he trusts, but not cow.

As in the previous section, our discussion focuses on a single, fixed object $k$ and we will not mention $k$ anymore. We write a positive belief as $v+$ and a negative belief as $v-$, where $v$ is a data value. An *explicit belief* can be positive $v+$, meaning that the user knows that the value is $v$, or can be a set of *negative beliefs* $v-, w-, \ldots$. We allow these sets to be infinite as long as they can be finitely represented, for example by a range predicate.

DEFINITION 3.1 (CONSISTENCY). *Two beliefs $b_1, b_2$ are conflicting ($b_1 \not\leftrightarrow b_2$) if they are either distinct positive beliefs $v+, w+$, or one is $v+$ and the other is $v-$. Otherwise, $b_1, b_2$ are consistent ($b_1 \leftrightarrow b_2$). A set of beliefs $B$ is called consistent if any two beliefs $b_1, b_2 \in B$ are consistent.*

DEFINITION 3.2 (PREFERRED UNION). *Given two consistent sets of beliefs $B_1, B_2$, their preferred union is:*

$$B_1 \vec{\cup} B_2 = B_1 \cup \{b_2 \mid b_2 \in B_2 . (\forall b_1 \in B_1 . b_1 \leftrightarrow b_2)\}$$

As in the previous section, our goal is to define, then compute all *implicit beliefs* based on the priority trust mappings. As we will see next, this raises both conceptual and computational challenges.

### 3.1 Three Paradigms

Consider the binary trust network in Fig. 6a. User $x_1$ defines a constraint resulting in a negative belief $b-$. Let's examine user $x_3$: she obviously adopts the explicit belief $a+$ from her preferred parent $x_2$. The question is: what should she do with the negative belief $b-$? Should her belief be $\{a+, b-\}$, or just $\{a+\}$? Clearly, once she believes the value $a+$, she has no more use for the constraint $b-$, since the purpose of the constraint was only to rule out $b+$ which is not under consideration at all here. This argument shows that she may well restrict her belief to $\{a+\}$. On the other hand, her decision may affect the users who trust her. Her immediate successor $x_5$ will reject $a+$, but the next user $x_7$ has the option of adopting $b+$ or not. The decision made by user $x_3$ affects whether $x_7$ can learn or not about the constraint $b-$ defined upstream. As this example shows, there are several choices in defining conflict-resolution in the presence of negative beliefs, even for graphs without cycles.

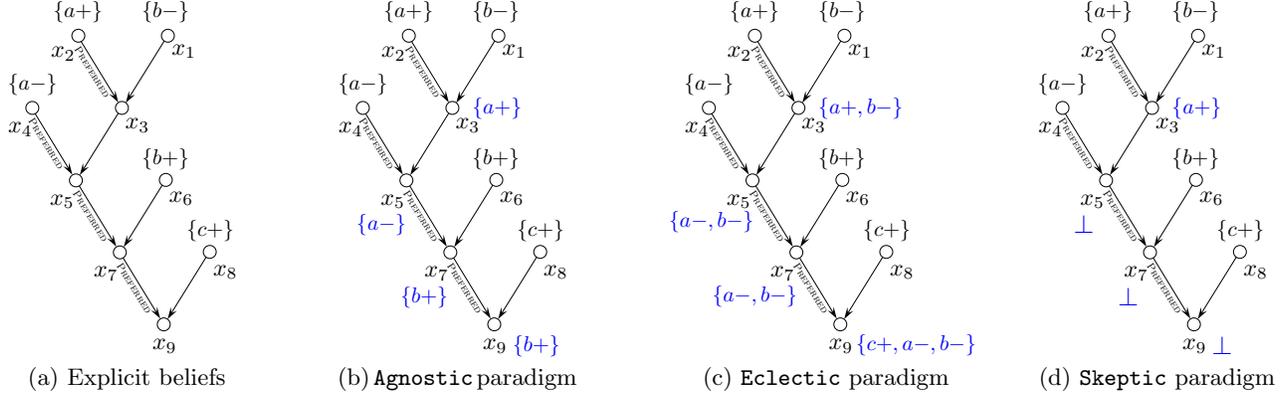

Figure 6: (a): An example binary trust network with explicit *positive and negative beliefs*. The edge from the *preferred* parent is labeled as such. (b-d): The three alternative paradigms lead to different entailed *implicit beliefs* at various nodes for the unique stable solution of the trust network.

We propose here three paradigms for trust network resolution in the presence of negative beliefs. A *paradigm* is formally defined as *a set of consistent sets of beliefs that are considered valid* or *in normal form*. We denote $\bot = \{v- \mid v \in D\}$ the set of *all* negative beliefs. Equivalently, $\bot$ is an inconsistent constraint that rejects any value.

**Agnostic.** The only valid belief sets in this paradigm are singleton positives $\{v+\}$ and sets of negatives $\{v-, w-, \ldots\}$. Once a user knows the value of an object, they do not want to know any constraints, even if they are consistent with this value. In the agnostic solution, the negative belief $b-$ is blocked by $x_3$ who believes only $a+$ (Fig. 6b).

*"An agnostic is a person who believes that nothing is known or can be known (...) beyond material phenomena."*

**Eclectic.** Any consistent set of beliefs is valid. In this paradigm, a user adopts all constraints that are consistent with a given value. Thus, $\{a+, b-, c-\}$ is a a valid set of beliefs. The eclectic solution is shown in Fig. 6c. Here $x_3$ accepts the constraint $b-$ in addition to $a+$. As a consequence, this constraint is communicated all the way to $x_7$, who now rejects $b+$.

*"An eclectic is a person who derives ideas, style, or taste from a broad and diverse range of sources."*

**Skeptic.** The valid sets of beliefs are the following: all sets with only negative beliefs, and all sets that contain exactly one positive belief and *all* negative beliefs consistent with it, i.e. they are of the form $\{v+\} \cup (\bot - \{v-\})$. Thus, when a user accepts a positive belief $v+$, she also adopts a constraint that rules out all other values. The skeptic solution is shown in Fig. 6d. In this paradigm the belief $a+$ "means" the set $\{a+, b-, c-, d-, \ldots\}$. When $x_5$ rejects $a+$, his belief becomes $\bot$. This propagates to $x_7$, who reject $b+$, similarly to the eclectic paradigm. However, at the next step $x_9$ rejects $c+$ too, hence $x_9$ does not belief any positive value (he believes $\bot$). This differs from the eclectic paradigm, where $x_9$ believes $c+$.

*"A skeptic is a person inclined to question or doubt all accepted opinions."*

The paradigm is chosen by the system administrator and applied to all users. The stable solutions to a trust network depend on the paradigm chosen. Before we can define the stable solutions, we need some technical definitions. Let $B$ be a consistent set of positive and/or negative beliefs. For each paradigm $\sigma \in \{\text{Agnostic}, \text{Eclectic}, \text{Skeptic}\}$ (abbreviated by $\{\text{A}, \text{E}, \text{S}\}$), the *normal form* $\text{Norm}_\sigma(B)$ is:

$$\text{Norm}_\text{A}(B) = \begin{cases} \{v+\} & \text{if } \exists v+ \in B \\ B & \text{otherwise} \end{cases}$$

$$\text{Norm}_\text{E}(B) = B$$

$$\text{Norm}_\text{S}(B) = \begin{cases} \{v+\} \cup (\bot - \{v-\}) & \text{if } \exists v+ \in B \\ B & \text{otherwise} \end{cases}$$

The *preferred union* specialized to the paradigm $\sigma$ is:

$$B_1 \vec{\cup}_\sigma B_2 = \text{Norm}_\sigma\bigl(\text{Norm}_\sigma(B_1) \vec{\cup} \text{Norm}_\sigma(B_2)\bigr) \quad (1)$$

For example:

$$\{a-\} \vec{\cup}_\text{A} \{b+\} = \{b+\}$$
$$\{a-\} \vec{\cup}_\text{E} \{b+\} = \{b+, a-\}$$
$$\{a-\} \vec{\cup}_\text{S} \{b+\} = \{b+, a-, c-, d-, \ldots\}$$
$$\{b-\} \vec{\cup}_\text{S} \{b+\} = \bot$$

We define next a stable solution for a binary trust network with constraints. We make the restriction that edges entering the same node have distinct priorities, thus we disallow ties. We discuss ties in Appendix B.

**Definition 3.3** (Stable solution w/ constraints). *Let $\sigma \in \{A, E, S\}$, and let $BTN = (U, E, B_0)$ be a binary trust network, where for all $x$, $B_0(x)$ is either a positive belief, or a set of negative beliefs, or the empty set. A stable solution is a function $B$ from users to sets of beliefs such that:*
(1) *If $x$ has a preferred parent $y$ and a non-preferred parent $z$, then $B(x) = B_0(x) \vec{\cup}_\sigma \bigl(B(y) \vec{\cup}_\sigma B(z)\bigr)$. If $x$ has only one parent $y$, then $B(x) = B_0(x) \vec{\cup}_\sigma B(y)$. If $x$ has no parent, then $B(x) = \text{Norm}_\sigma\bigl(B_0(x)\bigr)$.*
(2) *For every belief $b \in B(x)$ there exists a path $x_0 \to x_1 \to \ldots \to x_n = x$ such that $b \in \text{Norm}_\sigma\bigl(B_0(x_0)\bigr)$ and $b \in B(x_i)$ for all $i = 0, \ldots, n$.*

Consider a node $x$ and a positive belief $v+$. We say that $v+$ is *possible* if there exists a stable solution $B$ s.t. $v+ \in B(x)$. We say that $v+$ is *certain*, if $v+ \in b(x)$ for all stable solutions $B$. Our goal is to compute the possible and the certain positive beliefs under each of the three paradigms.

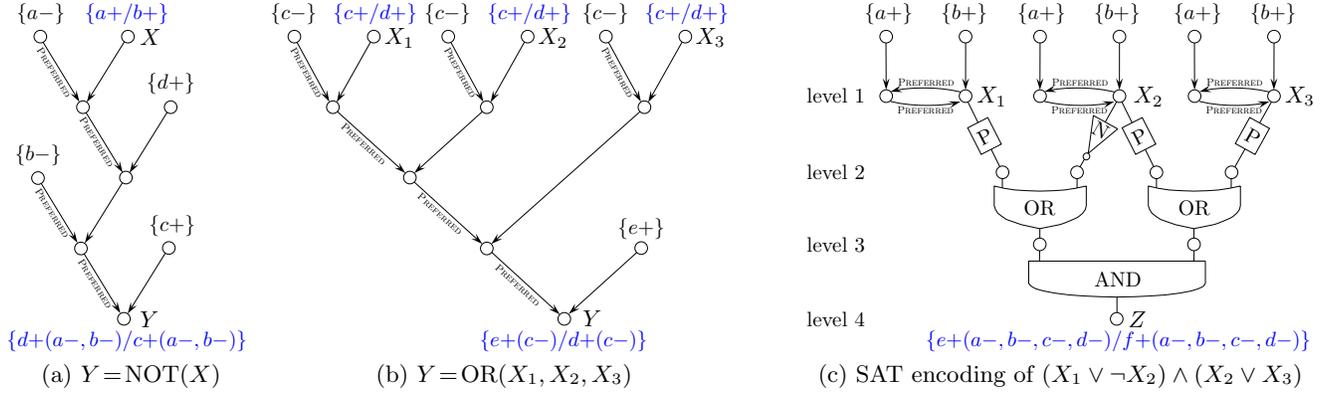

Figure 7: (a,b): Representation of NOT and OR gates for `Agnostic` and `Eclectic` paradigms. A PASS-THROUGH gate is similar to a NOT, an AND gate to an OR. (c): A full example of CNF. Notice that $0/1$ are encoded differently at different levels: the inputs at the top are $a/b$ for $0/1$, the output at $Z$ is $e/f$ for $0/1$.

## 3.2 The Complexities

We show next that the `Agnostic` and `Eclectic` paradigms are hard, whereas `Skeptic` is in PTIME.

THEOREM 3.4 (AGNOSTIC / ECLECTIC COMPLEXITIES). *Let $\sigma$ be `Agnostic` or `Eclectic`. Then the following hold: (a): Checking if a positive belief is possible at a given node $x$ in a binary trust network is NP-complete. (b): Checking if a positive belief is certain at a given node $x$ in a binary trust network is coNP-complete.*

PROOF SKETCH. The proof is a quite simple reduction from the CNF SAT problem, and we include it here because it highlights the inherent difficulties of coping with constraints in conflict resolution. The key technical step is that Boolean gates NOT, OR, and AND can be encoded as Trust Networks. Consider the NOT gate in Fig. 7a: The inputs $0/1$ are encoded as beliefs $a+/b+$, the outputs are encoded as $c+/d+$. When the input $X$ is $a+$ for 0, then the output $Y$ is $d+$ for 1. This is because $a+$ is blocked at the next level, hence $d+$ advances (the parentheses show the additional negative beliefs that are propagated in the `Eclectic` paradigm). Similarly, one can check that if $X$ is $b+$ for 1, the output $Y$ is $c+$ for 0. Thus, the network maps $a+/b+$ to $d+/c+$, hence it is a NOT. If we modify the network by switching $c$ with $d$, we obtain a PASS-THROUGH gate which maps $a+/b+$ to $c+/d+$. Next consider the ternary OR gate in Fig. 7b: If at least one of the three inputs is $d+$ for 1, $d+$ will propagate to the output. If all inputs are $c+$ then all positive beliefs are blocked, and the output is $e+$ for 0. Thus, the encoding of the output is $e+/d+$ for $0/1$. An AND gate is obtained similarly. By combining these gates, making sure that each level uses the same encoding of Boolean values as positive beliefs, we can represent an CNF expression where the last level encodes $0/1$ as $e+/f+$. We further use basic oscillators, as in Fig. 4b to generate inputs $0/1$. Fig. 7c illustrates the entire encoding of $(X_1 \vee \neg X_2) \wedge (X_2 \vee X_3)$. Notice the role of the PASS-THROUGH gates in ensuring that all values at the second level have the same encoding $c+/d+$. The reduction is completed by the observation that the CNF formula is satisfiable iff $f+ \in \text{poss}(Z)$, where $Z$ is the output node. This proves the first claim of the theorem. The second claim follows from the fact that checking non-satisfiability of CNF formulas is coNP-hard, and that the formula is unsatisfiable iff $e+ \in \text{cert}(Z)$. □

Given this negative result, it is somewhat surprising that the third paradigm `Skeptic` can be computed in PTIME. Algorithm 2 runs in $O(n^2)$ and computes a set $\text{repPoss}(x)$ for each node $x$, which is a representation of the set of possible values $\text{poss}(x)$ at $x$ as follows: If $\text{repPoss}(x)$ contains a positive value $v+$, then all other negative values are also possible. if $\text{repPoss}(x)$ contains $\bot$, then all negative values are also possible. In summary, the set of possible values $\text{poss}(x)$ represented by $\mathsf{R} := \text{repPoss}(x)$ is:

$$\text{poss}(x) = \{v- \mid v- \in \mathsf{R}\} \cup \{w- \mid \bot \in \mathsf{R}, w \in D\}$$
$$\cup \{v+ \mid v+ \in \mathsf{R}\} \cup \{w- \mid v+ \in \mathsf{R}, w \in D, w \neq v\}$$

If $\text{repPoss}(x)$ does not contain a positive value or $\bot$, we call it of Type 1. Otherwise, it is of Type 2.

We describe now the algorithm, which is a natural extension of Algorithm 1. The preprocessing phase (P) finds all nodes who either have explicit negative beliefs or whose preferred ancestors have explicit negative beliefs. For that it starts from nodes $x$ where the explicit belief has only negative values, and continues with all nodes whose preferred parent has only negative beliefs. These nodes correspond to the empty nodes in the earlier Algorithm 1, but unlike there, we cannot discard these nodes yet. The initialization phase (I) again creates the sets of closed and open nodes, and closes nodes with explicit positive beliefs. Step 1 (S1) of the main loop (M) traverses all preferred edges $z \to x$, and updates $\text{repPoss}(x)$ accordingly. Step 2 (S2) is slightly more involved and described next:

Here the algorithm finds a minimal connected component $S$ of open nodes (as before), but it cannot simply flood it with all positive values waiting to enter $S$. The reason is that some nodes in $S$ can be reached from closed nodes through preferred edges. These are precisely the edges we skipped in Step 1. Thus, some nodes in the component $S$ can be forced to accept negative beliefs, and these prevent a value $v+$ waiting to enter $S$ to reach the entire set $S$. Instead, we simply compute which subset of $S$ the value can reach, and add $v+$ to the set of possible values for those nodes. For all unreachable nodes, we add $\perp$ to the set of possible values, because in the `Skeptic` paradigm: $\{v-\}\vec{\cup}_\texttt{S}\{v+\} = \perp$.

THEOREM 3.5. (SKEPTIC RESOLUTION ALGORITHM) *Algorithm 2 runs in time $\mathcal{O}(n^2)$ and computes the set of possible values, for the `Skeptic` paradigm.*

## 3.3 Discussion

This section described how to handle constraints during conflict resolution. We represent constraints as negative beliefs and argue that they are an important feature in collaborative data sharing. The question we have studied is how trust mappings should handle constraints. Perhaps the most natural approach is to use the constraints only as filters for data values accepted from other users, but otherwise ignore them during reconciliation. This is what we called the `Agnostic` paradigm. The second natural approach is to simply propagate constraints together with data values, in what we called the `Eclectic` paradigm. However, we have shown that computing the possible values under both paradigms is NP-hard (and computing the certain values is co-NP hard), so we do not advocate their use in data reconciliation. Our third paradigm is, we believe, natural too: propagate constraints, but in addition associate to a data value a maximal constraint, which rules out any other data value. This paradigm, we have shown, is in PTIME, and the algorithm is a natural, yet somewhat detailed extension of Algorithm 1. We propose the `Skeptic` paradigm as the basis for conflict resolution in cooperative data sharing systems.

We note that, if no constraints exist in the system, then all three paradigms collapse to the simple semantics we discussed in Sect. 2.

The hardness of the `Agnostic` and `Eclectic` paradigms holds under the assumption that the network is cyclic: the proof used oscillators. If the network is a DAG, then all three paradigms can be computed in PTIME, by simply applying repeatedly the definition of preferred union (Eq. 1).

PROPOSITION 3.6 (ACYCLIC BTNS). *Let BTN be a binary trust network that is acyclic. Then for any of the three paradigms (`Agnostic`, `Eclectic`, `Skeptic`) the following hold: (a) there exists exactly one stable solution, and (b) that solution can be computed in PTIME.*

A puzzling question is why is the `Skeptic` paradigm in PTIME, while the other two are hard. It is easy to see that the Boolean gates in Fig. 7 no longer work under `Skeptic`, but we do not consider this a satisfactory explanation. While we cannot give an ultimate cause, we point out one interesting difference. The preferred union for `Skeptic` is *associative*, while it is not associative for either `Agnostic` nor `Eclectic`. For example, consider the two expressions $B_1 = \{a-\}\vec{\cup}_\sigma(\{a+\}\vec{\cup}_\sigma\{b+\})$, $B_2 = (\{a-\}\vec{\cup}_\sigma\{a+\})\vec{\cup}_\sigma\{b+\}$. For `Agnostic`, we have $B_2 = \{b+\}$, for `Eclectic` $B_2 = \{a-, b+\}$, while for both $B_1 = \{a-\}$. By contrast, one can show that $\vec{\cup}_\texttt{S}$ is associative. Associativity as a desirable property during data merging was pointed out in [14].

## 4. EXTENSION TO BULK PROCESSING

So far we have treated one object at a time. If several objects need to be updated, then the reconciliation algorithm needs to be run separately for each object. In this section, we show, that under certain conditions, the set of possible/certain values can be computed in bulk, for an entire

**Algorithm 2:** Skeptic Resolution Algorithm

**Input**: $BTN = (U, E, b_0)$
**Output**: repPoss($x$) for each node $x \in U$

foreach $x \in U$ do prefNeg($x$) $\leftarrow \emptyset$ and repPoss($x$) $\leftarrow \emptyset$
P  foreach *node $x$ with $\exists v- \in b_0(x)$* do
    | prefNeg($x$) $\leftarrow b_0(x)$

while $\exists$ *preferred edge $z \to x$ with $v- \in$ prefNeg($z$), $v+ \notin b_0(x)$* do
  | prefNeg($x$) $\leftarrow$ prefNeg($x$) $\cup$ prefNeg($z$)

I  closed $\leftarrow \emptyset$
foreach *node $x$ with $v+ \in b_0(x)$* do
  | repPoss($x$) $\leftarrow b_0(x)$
  | add $x$ to closed

open $\leftarrow U -$ closed
M  while open $\neq \emptyset$ do
S1    if $\exists$ *preferred edge $z \to x$ with $z \in$ closed, $x \in$ open and repPoss($x$) $= \emptyset$* then
      | repPoss($x$) $\leftarrow$ repPoss($z$)
      | move $x$ from open to closed
S2    else
      Let $SCC(\text{open})$ be the SCC graph constructed from the open nodes. Let $S = \{x_1, \ldots, x_n\}$ be a minimal SCC. Let $\{z_1, \ldots, z_k\}$ be all nodes in closed that have edges into $S$.
      forall $i \in \{1, \ldots, n\}$, $j \in \{1, \ldots, k\}$ do
        foreach $v+ \in$ repPoss($z_j$) do
          Let $S' = S - \{x \mid v- \in$ prefNeg($x$)$\}$
          if $\exists$ *path $z_j \to x_i$ in $S'$* then
            | repPoss($x_i$) $\leftarrow$ repPoss($x_i$) $\cup \{v+\}$
          else
            | repPoss($x_i$) $\leftarrow$ repPoss($x_i$) $\cup \{\perp\}$
        foreach $v- \in$ repPoss($z_j$) do
          | repPoss($x_i$) $\leftarrow$ repPoss($x_i$) $\cup \{v-\}$
      move all nodes of $S$ from open to closed

set of objects $k_1, k_2, \ldots, k_n$. We sketch here the approach and provide the details in Appendix B.

Let $TN_1, \ldots, TN_n$ be the trust networks for each of the $n$ objects. We make the following two assumptions:

(i) The set of trust mappings is the same for each object $k_i$, i.e. a user $x$ trusts a user $z$ *globally*, for all objects.

(ii) If a user has an explicit belief for an object $k_i$, then the user has an explicit belief for each of the objects.

Then it is possible to simply adapt both, Algorithm 1 and Algorithm 2 to bulk-compute the set of possible tuples through SQL queries. Let POSS(X,K,V) denote the a relation representing the possible values: an entry $(x, k, v)$ means that $v$ is a possible value for user $x$ and object $k$. Then, in step 1 of the modified algorithm, when traversing a preferred edge $z \to x$, we perform the following bulk insertion:

```
insert   into POSS
select   'x' AS X, t.K, t.V
from     POSS t
where    t.X = 'z'
```

In step 2, when 'flooding' a strongly connected component $SCC$ with the beliefs coming from the users $z_1, \ldots, z_k$ we perform the following bulk insertions for each $x_i \in SCC$:

```
insert   into POSS
select   distinct 'x_i' AS X, t.K, t.V
from     POSS t
where    t.X = 'z_1' or ... t.X = 'z_k'
```

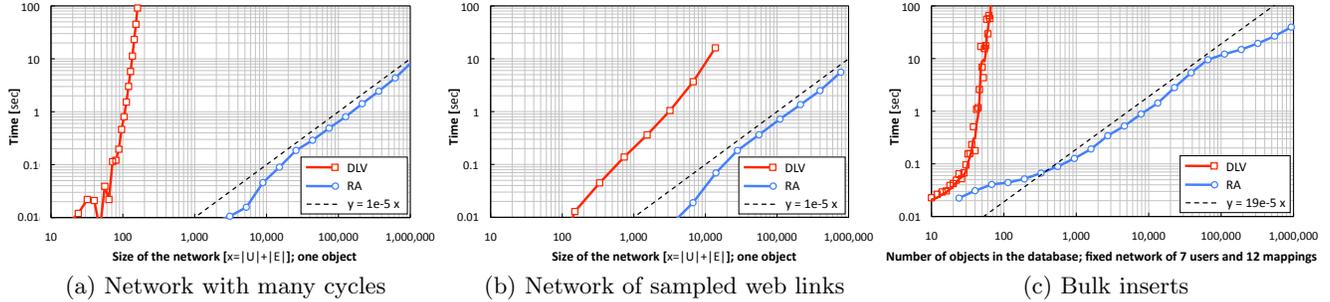

(a) Network with many cycles  (b) Network of sampled web links  (c) Bulk inserts

**Figure 8:** Experiments for various setups show quasi-linear scalability of our Resolution Algorithm (RA) in contrast to exponential scalability for solving the corresponding Logic Program with DLV.

For Algorithm 2 we need to modify some of the insert statements to insert the appropriate representation of $\perp$ rather than poss($x$).

## 5. EXPERIMENTS

We have conducted an experimental evaluation of our approach addressing the following questions: What is the observed runtime performance, both of the quadratic conflict resolution algorithm (RA) and of the bulk-update algorithm? And how does our approach compare to computing the stable models using an advanced liner program solver? For the latter we used the programming system DLV [12], which is widely considered the state-of-the-art implementation of logic programming with negation, and execute it under the brave query semantics[3]. All algorithms are implemented in Java and were run on a 2.2GHz dual core machine with 4GB of main memory. Execution times for RA are averaged over 20 trials.

We used two data sets in our experiments. The first is a synthetic data set, consisting of a network of many independent cycles. This network consists of several, disconnected 4-node clusters of the form from Example 2.6. In this network, one out of two users has an explicit belief. In our second data set, we modeled a trust network with as much expected real-life characteristics as possible. Real-world social structures are known to exhibit a power-law behavior, often referred to as scale-free. We used a large crawl of a top level domain of the World Wide Web and the link structure between sites which exhibits such behavior, with around 270k domains with 5.4M links. We identified domains with users and links with trust mappings, and assigned random priorities. We varied the size of this database by randomly sampling a fraction of all edges and include both start and end point in our sampled graph.

The results of our experiments are reported in Fig. 8. For the synthetic data set, as shown before, DLV runs in exponential time (Fig. 8a). In contrast, RA runs in almost linear time: thus, just increasing the number of cycles does not lead to a quadratic running time of our algorithm. We omit running times on very small data sets, because they showed high variance due to factors unrelated to the algorithm (Java memory management, L2 cache behavior). Figure 8b shows our results on the real data set. DLV performs better on this data set, since it contains fewer cycles, on average. However, our algorithm is still faster by several orders of magnitude and exhibits the same robust behavior as before. Overall, RA had an average running time of around 0.01 msec per user and trust relation on both data sets.

We note that the running time of our algorithm *is* quadratic in the worst case (we prove this in Appendix B), but the types of graphs that lead to quadratic behavior are complex and highly regular with nested cycles. We tested our algorithm on such kinds of graphs and measured running times of about 4 sec for a network size of 10,000.

Finally, Fig. 8c shows our result for testing the bulk conflict resolution algorithm from Sect. 4. We used a trust network with 7 users and 12 mappings, and assumed two users with explicit beliefs. Then we varied the number of objects in the database. For each object, we randomly chose these beliefs of the two users to be in conflict, or in agreement. Then we reconciled all users, on all objects. The database system was Microsoft SQL Server 2005, and we used JDBC for calls to the database. The data was stored in a relation POSS(X,K,V), as described in Sect. 4. Figure 8c shows linear data scalability of our algorithm[4]. The running time of DLV is exponential here too, because of the conflicts in half of the objects between the two users. In contrast, the running time of the bulk conflict resolution algorithm is independent of the number of conflicts.

## 6. RELATED WORK

Several systems have been recently described that adopt some kind of *conflict resolution* or *trust mapping* to model data sharing in a community of users [7, 9, 11, 16, 19]. These systems discuss techniques in the presence of various subsets of the features needed in conflict resolution (Fig. 3). Conflicts occur when the system allows key constraints; a conflict is simply a violation of the key constraint. Orchestra is the only system that considers both conflicts and priorities, and is closest in spirit to our setting. We share with Youtopia the use of certain/possible values, a concept originally introduced for incomplete databases [10].

Our solution has relevance to the research areas of *data integration* and *data exchange*. Similar to [19], we assume a fixed schema in this paper. In contrast, work on *peer data exchange* focuses on arbitrary tuple and equality generating dependencies between different database instances with varying schemas. However, it does not consider priorities amongst different mappings between peers and it is known

---

[3]command: "dlv.bin -brave input.txt query.txt"

[4]Data points below 100 msec seem to be dominated by the overhead of the Java-SQL server connection

that data complexity under this semantics is in coNP, and that it can be coNP-complete even for acyclic peer data exchange settings [5]. Those proposals whose semantics is tractable do not consider local integrity constraints (such as key violations) and hence do not model conflicts [11].

Using logic programs and stable models semantics as means to produce certain answers has been proposed before under the framework of *consistent query answering* (CQA) [1]. However, and as we have illustrated, this approach is generally not tractable; no polynomial time case is known for universal constraints with negation [3, Fig. 1]. As a consequence, previously proposed semantics that build on logic programming have coNP data complexity [2] and we are not aware of any proposed polynomial algorithm that deals with a principled solution to inconsistency resolution for key constraints and trust mappings.

## 7. CONCLUSIONS

This paper presents a principled approach to the problem of *conflict resolution* in a *community database* where users *trust* other users with varying priorities. Our semantics has a number of intuitive and desired properties such as independence of the update sequence and, hence, correctly models revoke operations and avoids transient effects. We give a PTIME algorithm that has quadratic worst case scalability in the size of the network. In our experiments, we show that it is actually linear on typical network structures. We also expand our techniques to constraints, and show, depending on the approach taken, conflict resolution can be either intractable, or computable in time that is quadratic in the number of users. In addition, we show how this algorithm can be easily adapted to include bulk processing.

**Acknowledgement.** We thank Raj Rao for background information on the Indus script and the anonymous reviewers for their helpful comments. This work was supported in part by NSF grants IIS-0513877, IIS-0713576, and IIS-0915054.

# APPENDIX

## A. PROOFS

PROOF SKETCH OF PROPOSITION 2.8. Let $TN = (U, E, b_0)$. We construct the binary network $BTN$ by adding some additional nodes to $U$, and re-defining the edges. First, for every node $x \in U$ with an explicit belief $v = b_0(x)$, we create a new node $x_0$ in $BTN$, define $b_0(x_0) = v$, and make $x_0$ the highest priority parent of $x$: thus, $x$ no longer has an explicit belief, but will get his belief from $x_0$, which is a root node in $BTN$. Second, let $(z_1, p_1, x), \ldots, (z_k, p_k, x)$ be all edges entering $x$, in increasing order of priority: $p_1 \leq \ldots \leq p_k$. We create $k-2$ new nodes $y_2, \ldots, y_{k-1}$, and further denote $y_1 = z_1$, $y_k = x$. We remove all edges entering $x$, and create edges $(y_{i-1}, 1, y_i)$ and $(z_i, p, y_i)$ for $i \in \{2, \ldots, k\}$, where $p = 2$ if $p_i > p_{i-1}$, and $p = 1$ if $p_i = p_{i-1}$. Thus, every node $y_i$, $i \in \{2, \ldots k\}$, has two parents $y_{i-1}$ and $z_i$, which are either both non-preferred (priorities 1, 1) or non-preferred and preferred (priorities 1, 2). We prove the following in Appendix B: (a) for every stable solution $b$ of $BTN$ its restriction to the nodes in $TN$ is a stable solution for $TN$, and (b) every stable solution $b$ of $TN$ can be extended to a stable solution of $BTN$. □

PROOF OF THEOREM 2.12. We call a set $U' \subseteq U$ *free* if (a) $U'$ contains all root nodes, and (b) every edge from a node in $U - U'$ to a node in $U'$ is a non-preferred edge. The intuition is that a preferred edge forces the target node to accept the value of the source node: $U'$ is free, if it is not forced to accept a value from $U - U'$.

We show by induction that at each step of the algorithm, closed is free. Initially it contains only root nodes, hence it is free. Step 1 extends closed with a new node $x$ by following a preferred edge: since preferred edges are unique for every node, there are no other preferred edges into $x$, hence closed is still free. In Step 2, closed is extended with an entire set $S$: but all edges entering $S$ are coming from closed (because we chose $S$ to be minimal), it continues to be free after extension. Hence, closed is free at each step of the algorithm.

LEMMA A.1 (FORWARD LEMMA). *Let $U'$ be a free set of $TN = (U, E, b_0)$, and let $E'$ be the set of edges where both endpoints are in $U'$. Let $b'$ be a stable solution for the subgraph $(U', E', b_0)$. Then $b'$ can be extended to the entire graph $TN$. More precisely, there exists a stable solution $b$ for $TN$ such that for all $x \in U'$, $b(x) = b'(x)$.*

PROOF SKETCH. We prove the statement by induction on the size of the set $U - U'$. When $U' = U$, then there is nothing to prove, since $b'$ already applies to the entire graph, so suppose $U' \subset U$. Case 1: there exists a preferred edge from a node $z \in U'$ to a node $x \in U - U'$. Extend $b'$ to $x$ by defining $b'(x) = b'(z)$; one can check directly from Definition 2.4 that this is a stable solution for $U' \cup \{x\}$, and thus we apply induction hypothesis to $U' \cup \{x\}$. Case 2: all edges from $U'$ to $U - U'$ are non-preferred. Let $S$ be a minimal connected component of $U - U'$, and let $z \to x$ be any edge[5] from $U'$ to $S$. Extend $b$ to the entire set $S$ by defining $b(y) = b(z)$ forall $y \in S$; one can check that this is a conflict-resolution solution for $U' \cup S$; apply induction hypothesis to the set $U' \cup S$. □

As a simple illustration of the power of Lemma A.1, recall that there exists logic programs without any stable model [4]; by contrast, every BTN has at least one stable solution. This follows from the Forward Lemma by letting $U'$ be the set of roots, and defining $b'(x) = b_0(x)$ for every root node $x$: the lemma ensures the existence of a stable solution.

We now use the lemma to prove the correctness of the algorithm, by showing soundness and completeness. *Soundness*: We show by induction on the number of steps that any value $v \in \text{poss}(x)$ is a possible value. Suppose the algorithm applies Step 1 to a preferred edge $z \to x$ and sets $\text{poss}(x) = \text{poss}(z)$. Let $v \in \text{poss}(z)$: by induction hypothesis there exists a stable solution $b$ s.t. $b(z) = v$. Then $b(x) = b(z)$, because the edge $z \to x$ is preferred. (We did not use the lemma at this step.) Suppose the algorithm applies Step 2 to the SCC $S$, and let $z \to x$ be a non-preferred edge entering $S$, and let $v \in \text{poss}(z)$. We will show that $v$ is a possible value for all nodes in $S$. By induction hypothesis there exists a stable solution $b$ s.t. $b(z) = v$. Define $U' = \text{closed} \cup S$: we have shown earlier that this is a free set. Define the stable solution $b'$ on $U'$ as follows: $b'$ is equal to $b$ on closed, and forall nodes $y \in S$, $b'(y) = b(z) = v$. One can check that $b'$ is indeed a conflict resolution on the set $U'$ by applying Definition 2.4 directly (here we use the fact that all preferred edges to nodes in $S$ are coming only from $S$). Apply the Forward Lemma to argue that it can be extended to a conflict resolution solution on the entire graph: this proves that $v$ is a possible value for all nodes in $S$. *Completeness*: We show by induction that for every node $x$ and every stable solution $b$, $b(x) \in \text{poss}(x)$. If $x$ enters closed after traversing a preferred edge, then the statement follows inductively from $x$'s parent, so assume $x$ is in a SCC $S$, and let $b$ be a stable solution. Traverse the lineage of $b(x)$ backwards, from $x$ to a root node, and consider the first node $z$ that is not in $S$. By induction hypothesis $b(z) \in \text{poss}(z)$, which implies that $b(x) = b(z) \in \text{poss}(x)$ (since $\text{poss}(x)$ is the union of all such $\text{poss}(z)$). Note that the completeness proof also implies that, if $\text{poss}(x)$ is a singleton set, then its unique element is a certain value: $\text{cert}(x) = \text{poss}(x)$.

---
[5]Since each node in the graph is reachable from a root node, and all root nodes are in $U'$, such an edge always exists.

PROOF OF PROPOSITION 2.13. We assume the constraint $(v, w) \in \text{poss}(x, y) \Leftrightarrow (w, v) \in \text{poss}(y, x)$ is maintained automatically, i.e. whenever we insert $(v, w)$ in $\text{poss}(x, y)$, we also insert $(w, v)$ into $\text{poss}(y, x)$. For each root node $x$ we initialize $\text{poss}(x, x) = \{(b_0(x), b_0(x))\}$. To compute $\text{poss}(x, y)$, the algorithm needs to be modified as follows. During Step 1, when we follow a preferred edge $z \to x$, we set for all $u \in \text{closed}$:

$$\text{poss}(u, x) = \text{poss}(u, z)$$
$$\text{poss}(z, x) = \{(v, v) \mid v \in \text{poss}(z)\}$$

During Step 2, when we compute the set of possible values for an SCC $S$, we do the following. Let $z_1, \ldots, z_k$ be all nodes in closed that have an edge entering $S$, and let $x_1, \ldots, x_k$ be their endpoints in $S$. Then:

$$\text{poss}(u, x) = \bigcup_{i=1,k} \text{poss}(u, z_i), \text{ where } u \in \text{closed}, x \in S$$
$$\text{poss}(x, y) = \bigcup_{i,j: \exists \text{disjoint paths}_{x_i \to x, x_j \to y} \text{ in } S'} \text{poss}(z_i, z_j)$$

We explain these two formulas next. From the proof of Theorem 2.12 we have seen that for every value $v \in \text{poss}(z_i)$ it is possible to flood the entire component $S$ with $v$. This proves the first line above: any pair of values possible for $(u, z_i)$ is also possible for $(u, x)$. For the second line, we need to check if, given $v \in \text{poss}(z_i)$ and $w \in \text{poss}(z_j)$, it is possible to assign $v$ to the node $x$ and $w$ to the node $y$. We proceed as follows. Let $S'$ be obtained from $S$ by collapsing all nodes that are connected via preferred edges. (In any stable solution, all collapsed nodes must have the same value). In the resulting graph, we check whether there exists two independent paths $x_i \to x$ and $x_j \to y$: this is solvable in PTIME using network-flow techniques [6, pp. 217]; in particular, if $x$ and $y$ are connected by a path of preferred edges, then they are the same node in $S'$ and there are no independent paths. If such a pairs of paths exists, then $x, y$ can take simultaneously values $v$ and $w$. To prove this, start with a stable solution $b$ that has $b(z_1) = v$ and $b(z_2) = w$, and partition $S'$ arbitrarily into two sets $S_1, S_2$ s.t. the first one contains the path $x_i \to x$ and the second one contains the path $x_j \to y$. All edges crossing between $S_1$ and $S_2$ are non-preferred (since we have collapsed preferred edges): hence we can extend the stable solution $b$ to $S$ by assigning the value $v$ to all nodes in $S_1$ and value $w$ to all nodes in $S_2$. This proves that the pair of values $(v, w)$ is possible for the nodes $x, y$. □

## B. APPENDIX FOR TECHNICAL REPORT

### B.1 Nomenclature

| | |
|---|---|
| $k$ | key representing an object |
| $D$ | domain, set of data values |
| $(k, v)$ | belief, key-value pair |
| $v, w, u$ | values $\in D$, beliefs if key is assumed fixed |
| $U$ | set of users |
| $x, y, z$ | users $\in U$ |
| $p$ | priority |
| $m$ | priority trust mapping $(z, p, x)$ with priority $p$ as specified by a user $x$ from a parent $z$, arrows point from parent to node $(z \to x)$ |
| $E$ | edges, set of priority trust mappings |
| $b_0$ | explicit beliefs, partial function $b_0 : U \to D$ |
| $b$ | beliefs $b : U \to D$, corresponds to one stable solution |
| $TN$ | a trust network $(U, E, b_0)$ |
| $BTN$ | a binary trust network $(U', E', b_0)$ |
| $r, s$ | variables used for values in Logic Programs (LPs) |
| $S$ | Strongly Connected Component (SCC) |
| $\mathsf{poss}(x)$ | possible belief for $x$, brave stable model semantics |
| $\mathsf{cert}(x)$ | certain belief for $x$, cautious stable model semantics |
| $B$ | sets of beliefs $B : U \to 2^D$ in section 3 |
| $a, \ldots, f$ | signed values in section 3 |
| $\vec{\cup}$ | Preferred Union between two sets of beliefs |
| $\sigma$ | Paradigm $\sigma \in \{\mathtt{Agnostic}, \mathtt{Eclectic}, \mathtt{Skeptic}\}$ also abbreviated by $\sigma \in \{\mathtt{A}, \mathtt{E}, \mathtt{S}\}$ |
| $b_1 \leftrightarrow b_2$ | consistent beliefs |
| $b_1 \not\leftrightarrow b_2$ | inconsistent beliefs |
| $\mathsf{repPoss}(x)$ | representation of $\mathsf{poss}(x)$ for $\mathtt{Skeptic}$ paradigm |
| $\mathbb{N}_j^k$ | set of natural numbers from $j$ to $k$, $\{j, \ldots, k\}$ |

### B.2 Stable model semantics

We give a short, self-contained review of logic programming and stable models and refer to [4] for details.

A *normal logic program* $P$ is a finite set of rules of the form

$$A_0 \leftarrow L_1, \ldots, L_m \qquad (m \geq 0)$$

where $A_0$ is an atom and each $L_i$ is a positive or negative literal, i.e. either an atom $A_i$ or a negated atom $\neg A_i$. The *Herbrand base* $B_P$ of a logic program $P$ is the set of all *ground atoms* that can be formed with predicates and constants from $P$. An *interpretation* $I$ of a logic program $P$ is any subset $I \subseteq B_P$ of its Herbrand base. A *model* of $P$ is an interpretation such that for each rule $A_0 \leftarrow L_1, \ldots, L_m$ in $P$, this interpretation satisfies the logical formula $\forall(\bar{x}) : (L_1 \wedge \cdots \wedge L_m) \Rightarrow A_0$ where $\bar{x}$ is a list of the variables in the rule. The ground instantiation of $P$, denoted by $ground(P)$ is the union of all rules from $P$ that can be obtained by replacing variables with constants from $P$ in every possible way.

A *definite logic program* $P$ is a normal logic program that consists of only Horn clauses, i.e. rules where each of the literals is positive. Each definite logic program has one unique *least model* $T_P^\infty$.

A normal logic program $P$ is *stratified* if there is an assignment $str(\cdot)$ of integers $0, 1, \ldots$ to the predicates $p$ in $P$ so that for each rule $r$ in $P$ the following holds: If $p$ is the predicate in the head of $r$ and $q$ the predicate in an $L_i$ from the body, then $str(p) \geq str(q)$ if $L_i$ is positive, and $str(p) > str(q)$ if $L_i$ is negative.

The *reduct* $P^I$ of a normal logic program $P$ by an interpretation $I$ is the set of ground clauses obtained from $ground(P)$ as follows:

(i) remove every clause $r$ with a negative literal $\neg A_i$ in the body if its atom $A_i \in I$;
(ii) then remove all negative literals from the remaining rules

Notice that $P^I$ is a definite logic program and has one least model. An interpretation $I$ is a *stable model* of $P$ if $I$ is the least model of $P^I$: $I = T_{P^I}^\infty$. Every stratified $P$ has a unique stable model. Unstratified rules increase complexity and can have none, one or several stable models. Stratified propositional logic programming with negation is P-complete (equivalently, stratified Datalog with negation is data complete for P). Propositional logic programming with negation under SMS is co-NP-complete (equivalently, Datalog with negation under SMS is data complete for co-NP). And, given a propositional normal logic program $P$, deciding whether it has a stable model NP-complete.

### B.3 Binarization of trust networks

Here we give the details for the binarization and the complete proof of Prop. 2.8 outlined in Appendix A.

PROOF OF PROPOSITION 2.8 (BTN EQUIVALENCE). Let $TN = (U, E, b_0)$ be a non-binary trust network. Consider a node $x \in U$ with $k > 2$ parents $z_1, \ldots, z_k$ and construct a modified trust network $TN' = (U', E', b_0)$ with $k - 2$ additional nodes $y_2, \ldots, y_{k-1}$ and some edges redefined as detailed in the transformation below, so that all $x, y_2, \ldots, y_{k-1}$ have exactly 2 parents. Denote $SG$ the *induced subgraph* over $x, z_1, \ldots, z_k$ of $TN$, and $SG'$ the induced subgraph over $x, y_2, \ldots, y_{k-1}, z_1, \ldots, z_k$ of $TN'$. We call $SG'$ the *binarization* of $SG$, and will show that binarization does not change the stable solutions for the original nodes of a trust network.

**Transformation.** We first describe binarization: Let $(z_1, p_1, x), \ldots, (z_k, p_k, x)$ be all edges entering $x$ in increasing order of priority $p_1 \leq \ldots \leq p_k$. Create new nodes $y_2, \ldots, y_{k-1}$, and further denote $y_1 := z_1$ and $y_k := x$. Remove all edges entering $x$, and for each $y_i$ ($i \in \mathbb{N}_2^{k-1}$) create two edges, each with either priority 2 (called *preferred edge* and depicted double-lined) or priority 1 (called *non-preferred edge* and depicted single-lined) as follows:

(a) If $p_1 = p_{i-1} = p_i$, then create $(y_{i-1}, 1, y_i)$ and $(z_i, 1, y_i)$ (see Fig. 9a).
(b) If $p_{i-1} < p_i = p_{i+1}$, then create $(z_i, 1, y_i)$ and $(z_{i+1}, 1, y_i)$ (see Fig. 9b).
(c) If $p_1 < p_{i-1} = p_i = p_{i+1}$, then create $(y_{i-1}, 1, y_i)$ and $(z_{i+1}, 1, y_i)$ (see Fig. 9c).
(d) If $p_1 < p_{i-1} = p_i < p_{i+1}$, then create $(y_{j-1}, 1, y_i)$ and $(y_{i-1}, 2, y_i)$, where $j$ is the minimum index for which $p_j = p_i$ (see Fig. 9d).
(e) If $p_{i-1} < p_i < p_{i+1}$, then create $(y_{i-1}, 1, y_i)$ and $(z_i, 2, y_i)$ (see Fig. 9e).

Node $x = y_k$ is treated just as another $y_i$, but restricted to cases (a), (d), or (e), i.e. as if $p_k < p_{k+1}$.

The resulting subgraph $SG'$ is a binary tree, shaped as a cascade, where every node $y_2, \ldots, y_k$ has exactly two parents (which are either preferred and non-preferred, or both non-preferred). Parents $z_j, \ldots, z_l$ with the same priority in $SG$ appear in a subtree whose root $y_{l-1}$ dominates the path from each parent with lower priority to $x$, but whose path to $x$ in turn is dominated by every parent with higher priority in the original $SG$ (see Fig. 10c). Consider, for example, node $x$ with 7 parents $z_1, \ldots, z_7$ with $p_1 = p_2 < p_3 = p_4 = p_5 < p_6 < p_7$ from Fig. 10a. Its binarization has additional nodes $y_2, \ldots, y_6$ and is shown in Fig. 10b. Notice that $z_3, z_4, z_5$

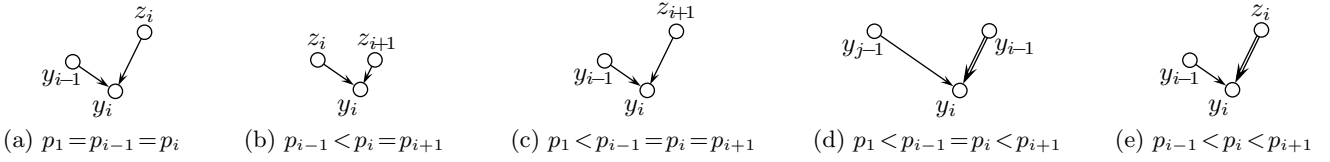

(a) $p_1\!=\!p_{i-1}\!=\!p_i$   (b) $p_{i-1}\!<\!p_i\!=\!p_{i+1}$   (c) $p_1\!<\!p_{i-1}\!=\!p_i\!=\!p_{i+1}$   (d) $p_1\!<\!p_{i-1}\!=\!p_i\!<\!p_{i+1}$   (e) $p_{i-1}\!<\!p_i\!<\!p_{i+1}$

Figure 9: Binarization: Each node $y_i$ with $i \in \mathbb{N}_2^k$ has edges to two parents with either different priorites (preferred and non-preferred) or equal priorities (both non-preferred).

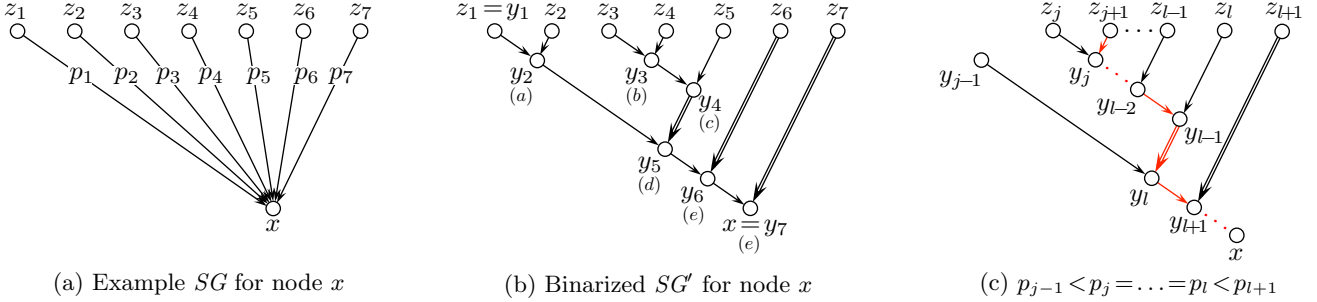

(a) Example $SG$ for node $x$   (b) Binarized $SG'$ for node $x$   (c) $p_{j-1}\!<\!p_j\!=\!\ldots\!=\!p_l\!<\!p_{l+1}$

Figure 10: Binarization: nodes with several incoming edges are cascaded into steps of binary resolution with preference or without preference. Subgraph $SG'$ in (b) is the binarization of $SG$ in (a) assuming $p_1\!=\!p_2\!<\!p_3\!=\!p_4\!=\!p_5\!<\!p_6\!<\!p_7$. Parentheses in (b) show which rules from Fig. 9 were applied. (c) Fragment of binarization for $x$'s parents $z_j, \ldots, z_l$ with equal priority: $p_{j-1}\!<\!p_j\!=\!\ldots\!=\!p_l\!<\!p_{l+1}$.

together with $y_3, y_4$ form a subtree with root $y_4$ that dominates all paths from $z_1$ or $z_2$ to $x$, but whose path to $x$ is dominated by the paths from $z_6$ or $z_7$ to $x$.

**Actual proof.** We next prove that (1) for every stable solution $b'$ of $TN'$ in which none of the nodes $y_2, \ldots, y_{k-1}$ has an explicit belief, its restriction to the nodes without $y_2, \ldots, y_{k-1}$ is a stable solution $b$ for $TN$, and (2) every stable solution $b$ of $TN$ can be extended to a stable solution $b'$ of $TN'$ by assigning appropriate beliefs from $z_1, \ldots, z_k$ to the additional nodes $y_2, \ldots, y_{k-1}$ in $SG'$. It follows that applying binarization for each of the original nodes with more than 2 parents in $TN$ leads to a binary trust network $BTN$ with exactly the same stable solutions over the original nodes in $TN$.

(1) $b' \to b$: We distinguish three cases:

(a): If $b'(x)$ is undefined in $TN'$, then $b'$ must be undefined as well for all $y_2, \ldots, y_{k-1}, z_1, \ldots, z_k$. Hence, $b$ is undefined for all $x, z_1, \ldots, z_k$ in $TN$, which is a stable solution.

(b): If $b'(x) = b'_0(x)$ for $TN'$, then $b(x) = b_0(x) = b'_0(x)$ for $TN$, which is a stable solution.

(c): If $b'(x)$ is defined, but $b'_0$ is undefined for all $x, y_2, \ldots, y_{k-1}$, then the lineage of $b'(x)$ must pass through at least one $z_{i^*}$ ($i^* \in \mathbb{N}_1^k$) in $TN'$. We show that keeping $b(x) = b'(x)$ and $b(z_i) = b'(z_i)$ for all $i \in \mathbb{N}_1^k$, but replacing the subgraph $SG'$ in $TN'$ with $SG$, is a stable solution with the lineage of $b(x)$ passing through the same $z_{i^*}$ in $SG$. We distinguish three subcases:

(i): First, assume that $z_{i^*}$ is part of a subtree such as the one with leaf nodes $z_j, \ldots, z_l$ in Fig. 10c (i.e. that $z_j, \ldots, z_l$ all have the same priority in the original $SG$). Then no node in $\{z_{l+1}, \ldots, z_k\}$ can have a belief with conflicting value as their path to $x$ would dominate that of $z_{i^*}$ to $x$. However, each node in $\{z_1, \ldots, z_l\} \setminus \{z_{i^*}\}$ may have conflicting beliefs as their path to $x$ cannot dominate the lineage of $b'(x)$. This solution corresponds exactly to the lineage for $b(x)$ in $SG$ as each node in $\{z_{l+1}, \ldots, z_k\}$ has a higher priority and each node in $\{z_1, \ldots, z_l\} \setminus \{z_{i^*}\}$ has a lower or equal priority.

(ii): Second, assume that $z_{i^*}$ is part of a subtree without preferred edges that includes $z_1$ such as the one with leaf nodes $z_1, z_2$ in Fig. 10b (i.e. that $\{z_1, \ldots, z_{i^*}, \ldots z_l\}$ all have the lowest priority in the original $SG$). Then no node in $\{z_{l+1}, \ldots, z_k\}$ can have a conflicting belief as their path to $x$ would dominate that of $z_{i^*}$ to $x$. This solution also holds for $b(x)$ in $SG$.

(iii): Third, assume that $z_{i^*}$ is connected to $y_{i^*}$ with a preferred edge such as $z_{l+1}$ in Fig. 10c (i.e. that nodes $\{z_{i^*+1}, \ldots, z_k\}$ have higher priorities, and $\{z_1, \ldots, z_{i^*-1}\}$ have lower priorities in the original $SG$). Then no node in $\{z_{i^*+1}, \ldots, z_k\}$ can have a conflicting belief, but each node in $\{z_1, \ldots, z_{i^*-1}\}$ can have. This solution also holds for $b(x)$ in $SG$.

(2) $b \to b'$: We again distinguish three cases:

(a): If $b(x)$ is undefined in $TN$, then $b$ must be undefined as well for all $z_1, \ldots, z_k$. Hence, $b'$ is undefined for all $x, y_2, \ldots, y_{k-1}, z_1, \ldots, z_k$ in $TN'$, which is a stable solution.

(b): If $b(x) = b_0(x)$ for $TN$, then $b'(x) = b'_0(x) = b_0(x)$ for $TN$. Each of the new nodes $y_2, \ldots, y_{k-1}$ gets consecutively assigned one value from its two parents according the following four simple rules: (i) if it has a preferred parent with a belief, it gets assigned this value; (ii) if it has two non-preferred parents with beliefs, it gets either of these values; (iii) if it has only one parent with a belief, it gets assigned this value; (iv) if it has no parent with a belief, it gets assigned no value. Note that the new nodes $y_2, \ldots, y_{k-1}$ are partially ordered, and consecutively assigning values to them with increasing index assures that the nodes are treated in this partial order (see Fig. 9). It follows that the resulting belief assignment $b'$ is a stable solution. Also note that while $TN'$ may have more total stable solutions for all variables including $y_2, \ldots, y_{k-1}$, the projection of these stable solutions onto the variables appearing in $TN$, i.e. without $y_2, \ldots, y_{k-1}$, is exactly the same stable solution $b$.

| | original $n$-clique | binarized network | relative factor for $\lim_{n\to\infty}$ |
|---|---|---|---|
| $\|U\|$ | $n$ | $n(n-2)$ | $n$ |
| $\|E\|$ | $n(n-1)$ | $2n(n-2)$ | $2$ |
| $\|E\|+\|U\|$ | $n^2$ | $3n(n-2)$ | $3$ |

**Figure 11: The maximal increase in size through binarization happens for an $n$-clique ($n \geq 4$), where number of edges increases by $< 2$ and number of edges plus nodes by $< 3$.**

(c): If $b(x)$ is defined, but $b_0(x)$ is undefined, then the lineage of $b(x)$ must pass through at least one $z_{i^*}$ ($i^* \in \mathbb{N}_1^k$) in $TN$. We show that keeping $b'(x) = b(x)$ and $b'(z_i) = b(z_i)$ for all $i \in \mathbb{N}_1^k$, while replacing the subgraph $SG$ in $TN$ with $SG'$ and assigning appropriate values to all nodes $y_2, \ldots, y_{k-1}$ is a stable solution with the lineage of $b'(x)$ passing through the same $z_{i^*}$ in $SG'$. This can be achieved as follows: Let $y_{j^*}$ be the single node that $z_{i^*}$ is connected to in $SG'$ (for example, this is $y_2$ for both $z_1$ and $z_2$, and $y_4$ for $z_5$ in Fig. 10b). Then assign all nodes $y_2, \ldots, y_{j^*-1}$ any belief according to the four simple rules given in (2b), and assign all nodes $y_{j^*}, \ldots, y_{k-1}$ the same value $b(x)$ in $TN'$. This is a stable solution, since the path $z_{i^*} \to y_{j^*} \to \ldots \to y_k = x$ cannot be dominated by any conflicting belief (other wise $b(x)$ would not be a stable solution in $TN$).

**Bounds.** We next give bounds for the increase in size of the network due to binarization. The induced subgraph $SG$ for a node with $k > 2$ parents gets binarized into a network $SG'$ with $k-2$ additional nodes. Hence, the number of edges in $SG'$ is $2(k-1)$, and the number of additional edges $2(k-1) - k = k-2$. Now consider a non-binarized network with $n$ nodes. Clearly, binarization introduces more additional nodes by having more edges between nodes. The worst case is a clique. Then, assuming $n \geq 4$, binarization adds $n-3$ new nodes for each node with $n-1$ parents, leading to $n(n-2)$ total nodes. The number of edges per node increases from $(n-1)$ to $2(n-2)$. Figure 11 compares the original number of nodes or edges with the binarized version, showing that the increase in number of edges is bounded by factor 2 and increase in number of edges and nodes by factor 3. □

**Remark.** Note that the cascading for the binarization has to be done with some care. Figure 12b and Fig. 12c show two possible approaches for binarization of the acyclic network in Fig. 12a (note the different orders of cascading). Whereas both binarized networks have the same stable solutions in the acyclic network, this symmetry does not hold anymore for cyclic networks: Fig. 12f is an incorrect binarization of the cyclic network Fig. 12c as $z_2$ and $z_1$ are symmetric in the binarized version, leading to incorrect stable solutions.

## B.4 Logic program equivalence

Here we prove Theorem 2.9. Instead of using unary predicates $U_x(v)$ and $C_{x,y}(v)$ for the Logic Program (LP) formulation as in Sect. 2.3, we use a binary predicate `poss(x,v)` and a ternary predicate `conf(x,y,v)`, respectively. Also, throughout this section, we use the actual syntax of DLV [12]

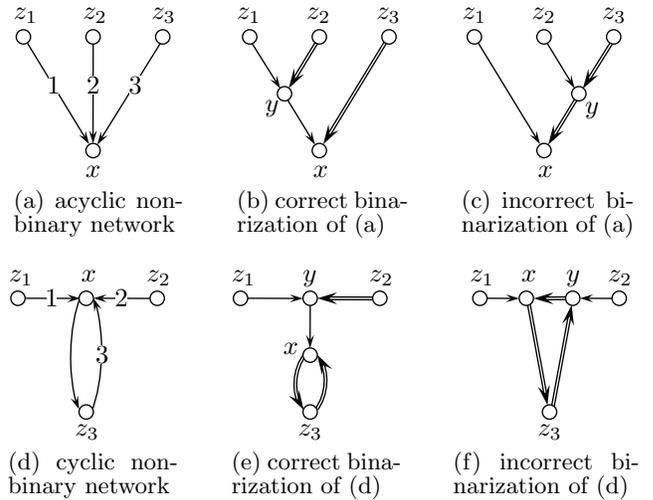

(a) acyclic non-binary network
(b) correct binarization of (a)
(c) incorrect binarization of (a)
(d) cyclic non-binary network
(e) correct binarization of (d)
(f) incorrect binarization of (d)

**Figure 12: Correct binarization of a trust network has to cascade from parents with lower priority to parents with higher priority as in (b) and (e).**

to write LPs that can be readily executed by DLV[6]. This translation of binary trust networks (BTNs) into LPs was used for the experiments in Sect. 5. Recall that the focus is on one single key.

In the syntax of DLV, constants begin with a lowercase letter (e.g. `x`, `z1` and `w`), whereas variables begin with an uppercase letter (e.g. `X`). Extensional database predicates (EDBs) such as

`poss(z1,v).`

encode explicit beliefs, here that node $z_1$ believes the value $v$. Intensional database predicates (IDBs) correspond to rules such as:

`poss(x,X) :- poss(z1,X).`

Executing both rules together as a logic program under the brave stable model semantics will deliver two result tuples `poss(z1,v)` and `poss(x,v)` with the second tuple being intensional and following from the LP. See Example B.1 for a detailed example.

PROOF THEOREM 2.9 (LOGIC PROGRAM EQUIVALENCE). We show that every stable solution to a BTN is a stable model to the associated LP, and vice versa. We give the transformation and the proof in one since the later follows directly the construction.

**Transformation and equivalence.** In a BTN, each node $x$ has either (a) no parent, (b) one parent, (c) two parents with different priorities, or (d) two parents with equal priorities. For cases (a) to (d), we assume that node $x$ has no explicit belief and consider a separate case (e) if $x$ has an explicit belief irrespective of the number of parents. For each of those 5 cases, we give a translation into a logic program (LP) and later show equivalence between the stable solutions of the BTN and the stable models of the LP.

(a) If node $x$ has neither explicit belief nor parents (Fig. 13a), then the LP contains no rule with head `poss(x,...)`.

---
[6] http://www.dbai.tuwien.ac.at/proj/dlv/

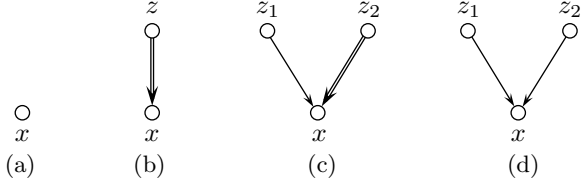

Figure 13: Four cases to be considered for translating the subgraph of a node without explicit beliefs and its parents into a logic program.

As consequence, all stable models have an empty set for values of $x$. Similarly, all stable solutions have no value for $x$. As a consequence, stable solutions and stable models are identical.

(b) If node $x$ has no explicit belief but one parent $z$ (Fig. 13b), then the LP for $x$ contains one rule

```
poss(x,X) :- poss(z,X).
```

This rule guarantees that, in all stable models, all possible values for $z$ are also possible values for $x$. This corresponds exactly to the stable solutions, hence they are identical.

(c) If node $x$ has no explicit belief but two parents with different priorities (Fig. 13c), then the LP consists of three rules:
```
poss(x,X)    :- poss(z2,X).
conf(x,z1,X) :- poss(z1,X), poss(x,Y), Y!=X.
poss(x,X)    :- poss(z1,X), not conf(x,z1,X).
```

The first rule guarantees that, as in case (b), in all stable models, all possible values from the preferred parent $z_2$ are also possible values for $x$. The second and third rule together guarantee that all possible values of the non-preferred parent $z_1$ that are not conflicting with a possible value of $x$ in a given stable model, are also possible values of $x$ in the same stable model. This rule corresponds exactly to the rule of stable solutions that the value of the non-preferred parent will only be used if it is not conflicting with the value of the preferred parent and, hence, the node $x$ itself. Hence, stable solutions and stable models are identical.

(d) If node $x$ has no explicit belief but two parents with equal priorities (Fig. 13d), then the LP consists of four rules:
```
conf(x,z1,X) :- poss(z1,X), poss(x,Y), Y!=X.
poss(x,X)    :- poss(z1,X), not conf(x,z1,X).
conf(x,z2,X) :- poss(z2,X), poss(x,Y), Y!=X.
poss(x,X)    :- poss(z2,X), not conf(x,z2,X).
```

Rule 1,2 and 3,4 are the same as rules 2,3 in case (c). Together, they guarantee that in each stable solution, node $x$ has exactly one of the values of their parents $z_1$ and $z_2$ if they have conflicting values. That means, if both parents together have possible values $\{v_1,\ldots,v_k\}$, then there are exactly $k$ different possible values for $x$. This corresponds exactly to the definition of a stable solution for the corresponding trust network. Hence, stable models and stable solutions are again identical.

(e) If node $x$ has an explicit belief $v$, then the only rule specified is an extensional rule

```
poss(x,v).
```

No other rule refers to `poss(x,...)`, hence all stable models contain $v$ as only value for $x$. Again stable solutions and stable models are identical.

Since all 5 above mutually exclusive cases exhaust all possible subgraphs for each node in a BTN, they together give a translation of a given BTN into a LP with the same stable models as the stable solutions for the BTN. Evaluating this LP under either the brave or the cautious stable model semantics give either the possible or certain values of the corresponding trust network. □

EXAMPLE B.1 (LOGIC PROGRAM WITH DLV). *Consider the binary trust network of Fig. 13c and assume $b_0(z_1) = v$ and $b_0(z_2) = w$. The corresponding logic program is*

```
poss(z1,v).
poss(z2,w).
poss(x,X)    :- poss(z2,X).
conf(x,z1,X) :- poss(z1,X), poss(x,Y), Y!=X.
poss(x,X)    :- poss(z1,X), not conf(x,z1,X).
```

*which we save as text file* `input.txt`. *In addition, we store the string "*`poss(X,U) ?`*" in a separate text file* `query.txt`, *which states a query for all tuples that appear in the intensional database predicate* `poss`. *We then execute DLV with the command* `dlv.bin -brave input.txt query.txt` *which issues the query over the logic program under the brave stable model semantics. DLV then returns three tuples* (`z1,v`), (`z2,w`), (`x,v`) *from which we learn that node $x$ has only one possible value $v$.*

*Similarly, the logic program for the binary trust network of Fig. 13d for $b_0(z_1) = v$ and $b_0(z_2) = w$ is*

```
poss(z1,v).
poss(z2,w).
conf(x,z1,X) :- poss(z1,X), poss(x,Y), Y!=X.
poss(x,X)    :- poss(z1,X), not conf(x,z1,X).
conf(x,z2,X) :- poss(z2,X), poss(x,Y), Y!=X.
poss(x,X)    :- poss(z2,X), not conf(x,z2,X).
```

*Executing this program results in four tuples* (`z1,v`), (`z2,w`), (`x,v`), (`x,w`) *from which we learn that node $x$ has two possible values $v$ and $w$.*

**Remark 1.** An alternative transformation is one in which case (e) is not treated differently. This can be achieved by creating a special node $x_0$ for each original node $x$ that represents a parent with higher priority than any other parent. If node $x$ has an explicit belief, this is then assigned as extensional predicate to the node $x_0$. The advantage of this alternative transformation is that all intensional rules are given by the trust network topology, independent of the explicit beliefs. The disadvantage is that the translation is larger since an extra parent node is created even for nodes without explicit beliefs, and the encoding of the parents is accounted for even if the node has an explicit belief.

**Remark 2.** Binarization is not essential for translating a TN into a LP. However, it simplifies the LP and reduces its size. Let $n$ be the number of nodes in *TN* and $k$ be the average number of parents of each node. The size of the resulting LP, i.e. the number of intensional rules, is $\mathcal{O}(nk)$ when the network is binarized. This follows from the fact that each node with $k$ parents gets binarized into maximum $k-1$ nodes, each of which has a maximum of four intensional rules in case (d). Hence, the size is $\mathcal{O}(nk)$.

However, if a trust network gets translated directly without binarization, the resulting LP is $\mathcal{O}(nk^2)$. The quadratic dependence on the number of parents results from the dependence of each parent on all parents with higher priority. This can be either encoded by $\mathcal{O}(k)$ number of rules per parent, or rules of size $\mathcal{O}(k)$. In addition, we for parents that share the priority with another parent, we need another rule specifies only one possible value per stable model. Instead of giving the formalities explicitly, we illustrate with the following two examples.

EXAMPLE B.2 (LP FOR NON-BINARY TN). *Consider the non-binary TN of Fig. 12a and assume no node has explicit values. The corresponding LP is*

```
poss(x,X)    :- poss(z3,X).
conf(x,z2,X):- poss(z2,X), poss(z3,Y), Y!=X.
poss(x,X)    :- poss(z2,X), not conf(x,z2,X).
conf(x,z1,X):- poss(z1,X), poss(z3,Y), Y!=X.
conf(x,z1,X):- poss(z1,X), poss(z2,Y), Y!=X.
poss(x,X)    :- poss(z1,X), not conf(x,z1,X).
```

*Note that 3 rules are necessary for parent $z_1$ which is possible blocked by 2 parents with higher priority.*

*Next consider the non-binary TN of Fig. 10a and assume no node has explicit values. The corresponding LP is*

```
poss(x,X)    :- poss(z7,X).
conf(x,z6,X):- poss(z6,X), poss(z7,Y), Y!=X.
poss(x,X)    :- poss(z6,X), not conf(x,z6,X).
conf(x,z5,X):- poss(z5,X), poss(z7,Y), Y!=X.
conf(x,z5,X):- poss(z5,X), poss(z6,Y), Y!=X.
conf(x,z5,X):- poss(z5,X), poss(x,Y),  Y!=X.
poss(x,X)    :- poss(z5,X), not conf(x,z5,X).
conf(x,z4,X):- poss(z4,X), poss(z7,Y), Y!=X.
conf(x,z4,X):- poss(z4,X), poss(z6,Y), Y!=X.
conf(x,z4,X):- poss(z4,X), poss(x,Y),  Y!=X.
poss(x,X)    :- poss(z4,X), not conf(x,z4,X).
conf(x,z3,X):- poss(z3,X), poss(z7,Y), Y!=X.
conf(x,z3,X):- poss(z3,X), poss(z6,Y), Y!=X.
conf(x,z3,X):- poss(z3,X), poss(x,Y),  Y!=X.
poss(x,X)    :- poss(z3,X), not conf(x,z3,X).
conf(x,z2,X):- poss(z2,X), poss(z7,Y), Y!=X.
conf(x,z2,X):- poss(z2,X), poss(z6,Y), Y!=X.
conf(x,z2,X):- poss(z2,X), poss(z5,Y), Y!=X.
conf(x,z2,X):- poss(z2,X), poss(z4,Y), Y!=X.
conf(x,z2,X):- poss(z2,X), poss(z3,Y), Y!=X.
conf(x,z2,X):- poss(z2,X), poss(x,Y),  Y!=X.
poss(x,X)    :- poss(z2,X), not conf(x,z2,X).
conf(x,z1,X):- poss(z1,X), poss(z7,Y), Y!=X.
conf(x,z1,X):- poss(z1,X), poss(z6,Y), Y!=X.
conf(x,z1,X):- poss(z1,X), poss(z5,Y), Y!=X.
conf(x,z1,X):- poss(z1,X), poss(z4,Y), Y!=X.
conf(x,z1,X):- poss(z1,X), poss(z3,Y), Y!=X.
conf(x,z1,X):- poss(z1,X), poss(x,Y),  Y!=X.
poss(x,X)    :- poss(z1,X), not conf(x,z1,X).
```

*Note that parent $z_2$ has 7 rules in total: 5 "blocking rules" for each of the 5 parents with higher priority, one "blocking rule" to allow only one belief (since $z_1$ and $z_2$ share priorities), and one "import rule".*

### B.5 Quadratic worst case scenario

We next show that the resolution algorithm (RA) Algorithm 1 is indeed $\mathcal{O}(n^2)$ in the worst case where $n$ is the number of nodes. Concretely, we show that RA runs in time quadratic of the size of the parameterized graph in Fig. 14a. Taking $k$ as parameter, the graph has $n(k) = 5 + 6k$ nodes and $m(k) = 5 + 10k$ edges, i.e. its size is linear in $k$. Starting at nodes $z_1$ and $z_2$, RA calculates $k$ times the connected components of the remaining nodes when arriving at nodes $y_{j,1}$ and $y_{j,4}$. Hence, RA runs in quadratic time of the size of this network. Figure 14b shows this progression and Fig. 15 shows experimental evidence.

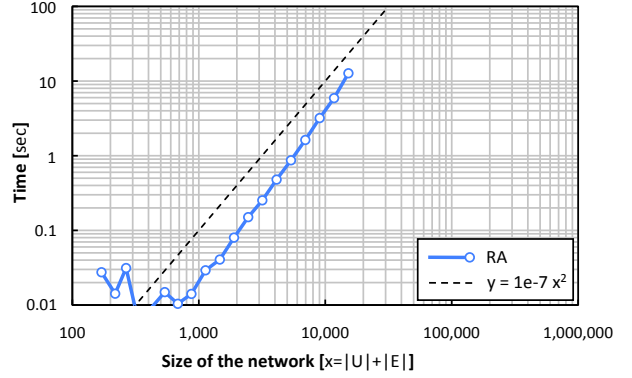

**Figure 15:** The RA (Algorithm 1) runs in quadratic time on the parameterized trust network of Fig. 14a.

**Remark.** Note that we only showed that *our algorithm* is $\mathcal{O}(n^2)$, not the problem. We doubt, but have not proven, that there is an algorithm that can run provably faster on any trust network and without prior knowledge of the topology of the graph. On the other hand, it seems very unlikely that any reasonable social network shows this regularity of "nested connected components" as illustrated with Fig. 14a that is required for quadratic scalability.

### B.6 Details on the hardness proof

PROOF THEOREM 3.4. The proof is reduction from the CNF SAT problem: given a formula $\Phi$ over Boolean variables $X_1, \ldots, X_n$ in CNF, determine whether there is a truth assignment of the variables so that the formula evaluates to true. We construct a BTN where all explicit beliefs (positive or negative) are over six data values $a, b, c, d, e, f$, and where $f+$ is a possible value for a distinguished node $Z$ iff there exists a satisfying assignment for $\Phi$ (Fig. 16f). The key technical step is that Boolean gates NOT, OR, and AND can be encoded as BTNs, and each of those gates changes the alphabet. Hence, we construct the BTN representing $\Phi$ level by level, and at each level we change the encoding of 1 (= true) and 0 (= false) (Fig. 17).

|         | 1 = true              | 0 = false             |
|---------|-----------------------|-----------------------|
| Level 1 | $b+$                  | $a+$                  |
| Level 2 | $d+(a-,b-)$           | $c+(a-,b-)$           |
| Level 3 | $d+(a-,b-,c-)$        | $e+(a-,b-,c-)$        |
| Level 4 | $f+(a-,b-,c-,d-)$     | $e+(a-,b-,c-,d-)$     |

**Figure 17:** Encoding of truth value at 4 levels of the BTN. Values in parenthesis show additional negative beliefs for the `Eclectic` paradigm.

We start at level 1 with a $b+/a+$ oscillator for each variable $X_i$ in $\Phi$ (Fig. 16a). That is, there is a node $X_i$ for each $i \in \{1, \ldots, n\}$ whose possible values are $b+$ and $a+$, representing 1 and 0, respectively.

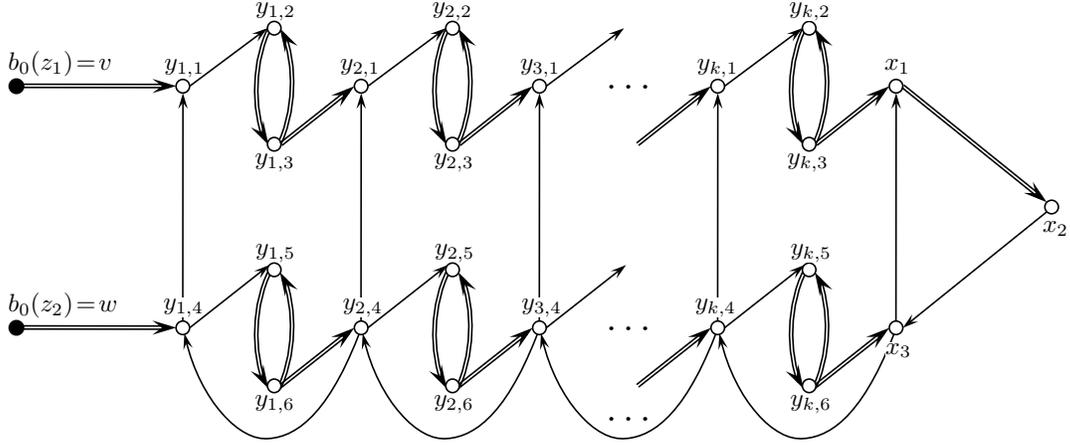

(a) BTN with size $\mathcal{O}(k)$

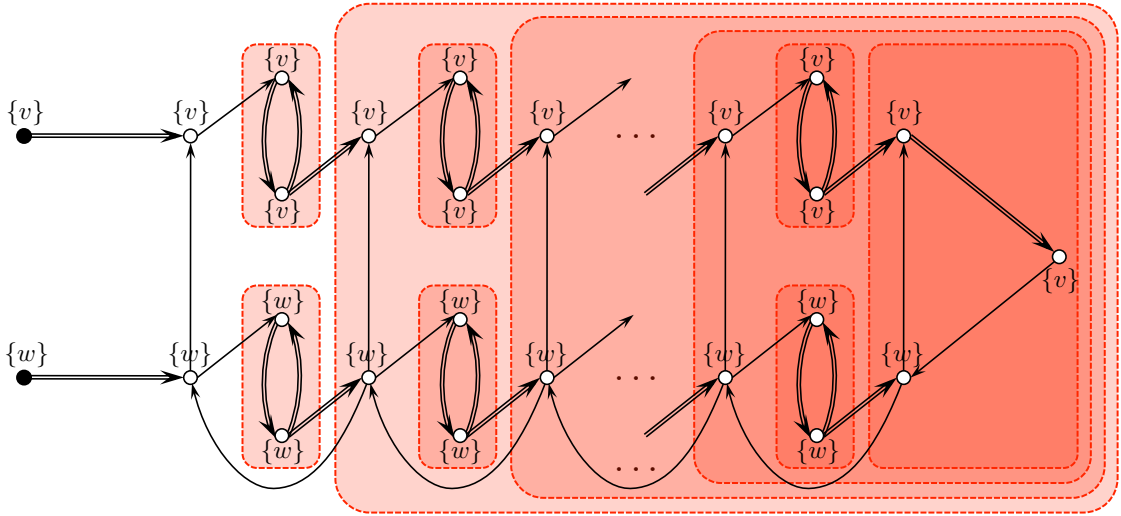

(b) Nested SSCs

Figure 14: (a) BTN which grows linear in size with parameter $k$, and for which RA needs $\mathcal{O}(k^2)$ time to determine value assignments. The reason is the "nestedness" of the connected components, which requires repeated application of Tarjan's algorithm (b).

At level 2, we construct a maximum of two new nodes $X_i^1$ and $X_i^0$ for each variable $X_i$, which represent $X_i$ and $\neg X_i$, respectively, and whose possible beliefs are $d+/c+$. $X_i^0$ is obtained from a NOT gate, $X_i^1$ is obtained from a PASS-THROUGH gate. Whereas the inputs of a NOT gate are $b+/a+$ representing 1/0, the outputs are $c+/d+$ representing 0/1. Thus the gate performs a negation. Consider the NOT gate in Fig. 16b: The inputs 1/0 are encoded as beliefs $b+/a+$, the outputs are encoded as $c+/d+$. When the input $X_i$ is $b+$ for 1, then the output $X_i^0$ is $c+$ for 0. This is because $b+$ is blocked at the node with preferred parent $\{b-\}$, hence $c+$ advances to the output (the parentheses show the additional negative beliefs that are propagated in the Eclectic paradigm). Similarly, one can check that if $X_i$ is $a+$ for 0, the output $X_i^0$ is $d+$ for 1. Thus, the network maps $b+/a+$ to $c+/d+$, hence it is a NOT. If we modify the network by switching $c$ with $d$, we obtain a PASS-THROUGH gate which maps $b+/a+$ to $d+/c+$ (see Fig. 16c).

Next, we apply an OR-gate for each clause in CNF expression $\Phi$ and obtain level 3 nodes, where the 1/0 encoding is $d+/e+$. Consider the ternary OR gate in Fig. 16d: if at least one the three inputs is $d+$ for 1, it will propagate to the output. If all inputs are $c+$ then all positive beliefs are blocked, and the output is $e+$ for 0. Thus, the encoding of the output is $d+/e+$ for 1/0.

Finally, we construct one AND-gate which leads to the single node $Z$ at level 4, and whose possible positive values are $f+/e+$, representing 1/0. Figure 16e illustrates a 3-way AND gate, which clearly generalizes to a $k$-way AND gate for any $k$.

By combining these gates, making sure that each level uses the same encoding of Boolean values as positive beliefs, we can represent an CNF expression where the last level encodes 1/0 as $f+/e+$. Figure 16 illustrates the entire encoding of a SAT $(X_1 \vee \neg X_2) \wedge (X_2 \vee X_3)$, which generalizes to any CNF SAT. Notice the role of the PASS-THROUGH gates (the rectangles) in ensuring that all values at the second level have the same encoding $c+/d+$.

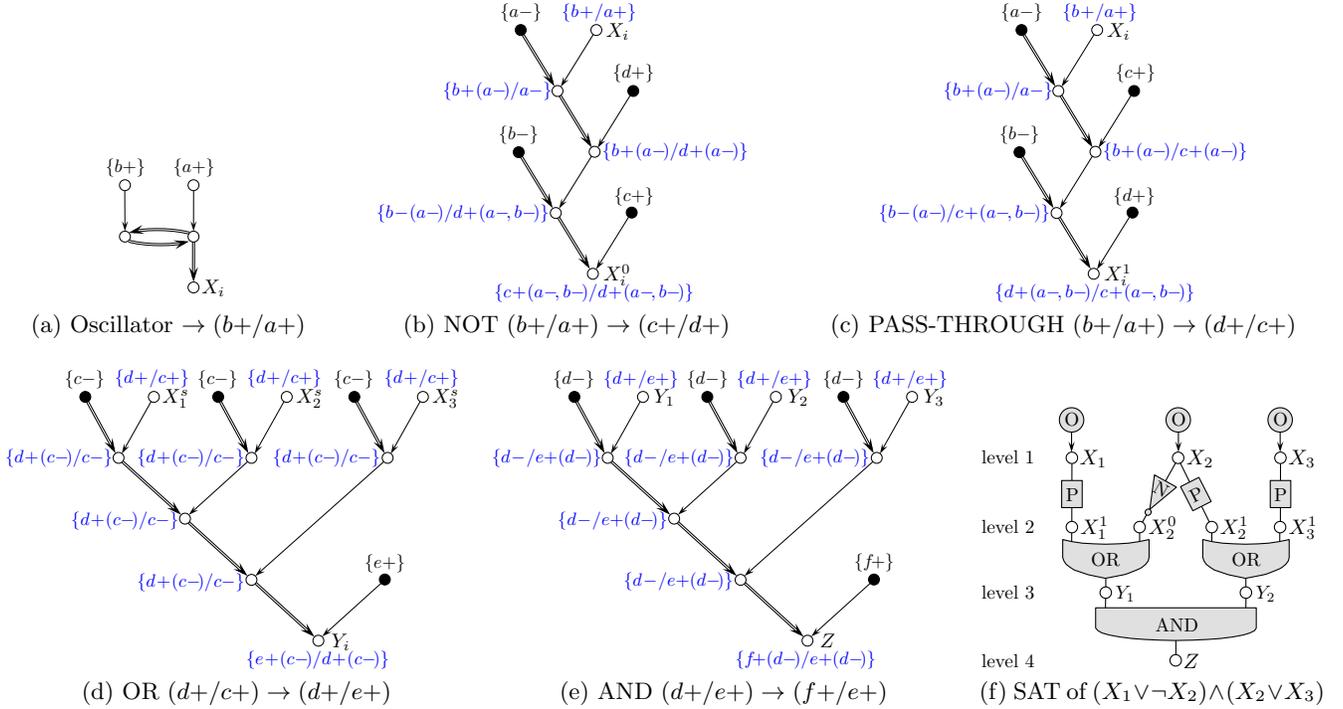

Figure 16: Oscillator (a), Pass-through gate (b), NOT gate (c), OR gate (d), AND gate (e), and example CNF SAT encoding of $(X_1 \vee \neg X_2) \wedge (X_2 \vee X_3)$.

| Case | Condition of repPoss($x$) | poss($x$) | cert($x$) |
|---|---|---|---|
| 1 | repPoss($x$) has only negative beliefs | repPoss($x$) | repPoss($x$) |
| 2 | repPoss($x$) contains $\bot$ and negative beliefs | $\bot$ | $\bot$ |
| 3 | $v+$ is the sole positive belief in repPoss($x$) and $v- \notin$ repPoss($x$) | $\{v+\} \cup (\bot - \{v-\})$ | $\{v+\} \cup (\bot - \{v-\})$ |
| 4 | $v+$ is the sole positive belief in repPoss($x$) and $v- \in$ repPoss($x$) | $\{v+\} \cup \bot$ | $\bot - \{v-\}$ |
| 5 | $v_1+, \ldots, v_k+$ are all positive beliefs in repPoss($x$) and $k \geq 2$ | $\{v_1+, \ldots, v_k+\} \cup \bot$ | $\bot - \{v_1-, \ldots, v_k-\}$ |

Figure 18: Encoding of positive *and negative* possible and certain values in repPoss($x$) after Algorithm 2 terminates. Projecting away all non-positive values gives *positive* possible and certain beliefs.

The reduction is completed by the observation that the CNF formula is satisfiable iff $f+ \in$ poss($Z$), where $Z$ is the output node. This proves the first claim of the theorem. The second claim follows from the fact that checking non-satisfiability of CNF formulas is coNP-hard, and that the formula is unsatisfiable iff $e+ \in$ cert($Z$). □

## B.7 Details on the Skeptic Resolution Algorithm

The correctness proof extends the proof of Algorithm 1. The only change from before is that if there is a preferred edge $z \rightarrow x$, and $z$ is closed but repPoss($z$) is of Type 1, then the algorithm cannot yet close the node $x$. This is because $x$ can further receive some positive values through non-preferred edges.

At the end, each node may contain positive beliefs $v+$, negative beliefs $w-$ and $\bot$. Figure 18 illustrates five different cases for repPoss($x$) and the set of possible and certain values that it represents after the algorithm terminates. Note that this set represents both positive *and negative* beliefs. From poss($x$) and cert($x$), one can easily restrict to just the positive possible and certain beliefs.

## B.8 Acyclic trust networks

PROOF PROP. 3.6. When a trust network is acyclic, then there is a partial order $\preceq$ between all nodes in the network. Visiting each node in any linear extension $\leq$ of this partial order, i.e. any total order that is consistent with the partial order, a node is always visited after all of its parents. Furthermore, if both preferred and non-preferred parent of a node are closed then it's value can directly be calculated by applying the preferred union. Hence, all values can be determined in $\mathcal{O}(n)$. □

## B.9 Conflict resolution with constraints and ties

If we allow both constraints and ties between between parents of a node, then the definition of a stable solution (Def. 3.3) has to be adapted. We generalize the random tie-breaking of Def. 2.4 by imposing a random order between non-preferred parents, then applying the overriding union in

exactly this order:

DEFINITION B.3 (STABLE SOLUTION W/ CONSTRAINTS 2).
Let $\sigma \in \{A, E, S\}$, and let $BTN = (U, E, B_0)$ be a binary trust network, where for all $x$, $B_0(x)$ is either a positive belief, or a set of negative beliefs, or the empty set. A stable solution is a function $B$ from users to sets of beliefs such that:
(1) If $x$ has a preferred parent $y$ and a non-preferred parent $z$, then $B(x) = B_0(x) \vec{\cup}_\sigma \big(B(y)\vec{\cup}_\sigma B(z)\big)$.
(2) If $x$ has two non-preferred parents $y$ and $z$, then either $B(x) = B_0(x) \vec{\cup}_\sigma \big(B(y)\vec{\cup}_\sigma B(z)\big)$, or $B(x) = B_0(x) \vec{\cup}_\sigma \big(B(z)\vec{\cup}_\sigma B(x)\big)$.
(3) If $x$ has only one parent $y$, then $B(x) = B_0(x)\vec{\cup}_\sigma B(y)$.
(4) If $x$ has no parent, then $B(x) = Norm_\sigma\big(B_0(x)\big)$.
(5) For every belief $b \in B(x)$ there exists a path $x_0 \to x_1 \to \ldots \to x_n = x$ such that $b \in Norm_\sigma\big(B_0(x_0)\big)$ and $b \in B(x_i)$ for all $i = 0, \ldots, n$.

Note that the order in which the overriding union is applied corresponds to the cascading for binarization described in Fig. 12. This allows to define the stable solutions for non-binary trust networks as well, so that the logic of binarization of Sect. B.3 applies. Hence, again each non-binary trust network can be translated into an equivalent binary trust network.

DEFINITION B.4 (STABLE SOLUTION W/ CONSTRAINTS 3).
Let $\sigma \in \{A, E, S\}$, and let $TN = (U, E, B_0)$ be a binary trust network, where for all $x$, $B_0(x)$ is either a positive belief, or a set of negative beliefs, or the empty set. A stable solution is a function $B$ from users to sets of beliefs such that for each node $x$ in the network the following holds:
(1) Let $y_i$ with $i \in \{0, \ldots, k\}$ be the parents of $x$ with their priorities defining a total preorder $\lesssim$. Then there exists a linear extension of $<$ of the total preorder with $y_1 < y_2 < \cdots < y_k$ so that $B(x) = Norm_\sigma\big(B_0(x)\big) \vec{\cup}_\sigma \big(B(y_k)\vec{\cup}_\sigma \big(B(y_{k-1}) \cdots \vec{\cup}_\sigma B(y_1)\big)\big)$.
(2) For every belief $b \in B(x)$ there exists a path $x_0 \to x_1 \to \ldots \to x_n = x$ such that $b \in Norm_\sigma\big(B_0(x_0)\big)$ and $b \in B(x_i)$ for all $i = 0, \ldots, n$.

## B.10 Details on Bulk Inserts

Remember the logic program formulation from Sect. B.4. We extend this schema with one more attribute K for the key of the value. Hence, we have one single relation POSS(X,K,V) that stores the appropriate value V for each user X and each key K.

We make the two following key assumptions:
(i) The set of trust mappings is the same for each object $k_i$, i.e. a user $x$ trusts a user $z$ globally, for all objects.
(ii) If a user has an explicit belief for an object $k_i$, then the user has an explicit belief for each of the objects.

These two assumptions imply that the resolution algorithm resolves nodes in the same order *across all keys*. As a consequence, we need to run the resolution algorithm on the network only once to determine the sequence of resolution steps, then apply those across all keys at the same time.

Whenever step 1 (S1) is applied, then all values from the preferred parent $z$ are copied to the child $x$ ($z \to x$):

```
insert   into POSS
select   'x' AS X, t.K, t.V
from     POSS t
where    t.X = 'z'
```

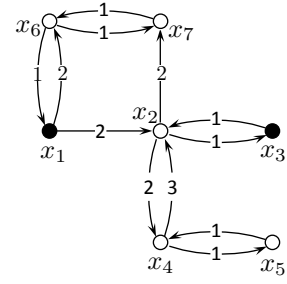

Figure 19: Non-binary network used for the bulk insert experiments. Dark nodes show users with explicit beliefs.

For step 2 (S2), the algorithm determines a minimal SCC of the open nodes, then calculates $\mathsf{poss}(x_i)$, for all nodes $x_i \in SCC$ as the union of all possible values of all parents $z_1, \ldots z_k$ of nodes in $SCC$:

```
insert   into POSS
select   distinct 'x_i' AS X, t.K, t.V
from     POSS t
where    t.X = 'z_1' or ... t.X = 'z_k'
```

Here the union is replaced by the logical or. Since we include the keys at each step, the values across different keys are kept separate.

Note that every node is visited only once. As a consequence, the number of SQL statements is linear in the size of the network. Determining the right sequence is still quadratic in the worst case, but is performed only once for the network. Hence, the resulting algorithm is quadratic in the size of the network, but linear in the number of objects.

Also note that our two assumptions are arguably strong: If they do not hold, then the resolution across different keys cannot be combined, and the number of SQL queries can be linear in the number of nodes *and* number of keys. It remains to be seen whether clever mechanism can still exploit common resolution patterns across different keys, in the general case.

Finally, Fig. 19 shows the network that was used for the experiments from Fig. 8c.